\documentclass[12pt,paper]{JHEP}
\usepackage{epsfig}
\newcommand{\be}{\begin{equation}}
\newcommand{\ee}{\end{equation}}
\newcommand{\ba}{\begin{eqnarray}}
\newcommand{\ea}{\end{eqnarray}}
\newcommand{\dis}{\displaystyle}

\title{The $\Delta I=1/2$ Rule in the Chiral Limit}
\author{Johan Bijnens\\Department of Theoretical Physics 2, University of Lund,
\\S\"olvegatan 14A, S 22362 Lund, Sweden}
\author{Joaquim Prades\\Departamento de F\'{\i}sica Te\'orica y del Cosmos,
Universidad
de Granada,\\Campus de Fuente Nueva, E-18002 Granada, Spain}
\abstract{We discuss the matching between long-distance and short-distance 
at  next-to-leading  in $1/N_c$
and show how the scheme-dependence from the two-loop renormalization group
running can be treated. We then use this method to
study the three $O(p^2)$ terms contributing to non-leptonic
kaon decays, namely the
usual octet and 27-plet derivative terms as well as the weak mass term
using the Extended Nambu--Jona-Lasinio model as the low energy approximation.
We also discuss subtleties in the momentum routing in the low energy theory
and a problem in separating factorizable and non-factorizable 
contributions from the $Q_6$ operator in the chiral limit.
We update our earlier results on the $B_K$ parameter  as well.}
\keywords{Weak Decay, Kaon Physics, Chiral Lagrangians, 1/N Expansion}
\preprint{{LU TP 98--26}\\
{UG--FT--94/98}\\{hep-ph/9811472}\\Revised January 99}

\begin{document}
\section{Introduction}

The $\Delta I=1/2$ rule in kaon decays has been the subject of
very many efforts at understanding it, see \cite{kreview} for a review.
We briefly discuss it and
a short history of attempts to understand it in Section \ref{deltaI}.
In this paper we attempt to put together various approaches
that have been done before. The short-distance effects are now known to
two-loops and the extended Nambu--Jona-Lasinio Model enhanced by using
Chiral Perturbation Theory whenever possible provides a reasonable
basis for the long-distance description of hadronic interactions needed.
We put the two together in a way that treats the scheme dependence
correctly. The underlying method, reproducing the results of the short-distance
running by an effective theory of exchanges of heavy bosons,
which we call $X$-bosons, is discussed in Section \ref{Xboson}.
The low energy model is shortly discussed in Section \ref{ENJL}.
In Section \ref{twopdef} we recall the definitions of the off-shell two-point
functions that we use here to determine the weak non-leptonic couplings.
The method here is basically to calculate these two-point functions
to next-to-leading order in $1/N_c$, but to all orders in the terms
enhanced by large logarithms involving $M_W$. We then compare with
the Chiral Perturbation Theory (CHPT) calculations of the same quantity
and in the end we calculate the relevant physical matrix elements using
CHPT.

In Section \ref{BKsect} we update our earlier results for $B_K$\cite{BPBK}.
Here we discuss in some detail the routing issue in Section \ref{routing},
which is rather non-trivial in the presence of neutral $X$-bosons whose
{\em direction} is not obvious. This also explains the discrepancies
of the results for very low $\mu$ in the chiral limit of \cite{BPBK}
and the results of \cite{FG95}. We give therefore updated numbers
and expressions for the main results of \cite{BPBK} here.

Section \ref{deltaS1longdistance} contains the same discussion
but for the $\Delta S=1$ operators $Q_1$ to $Q_6$. The current$\times$current
operators $Q_1$, $Q_2$, and\footnote{We use $Q_4=Q_2-Q_1+Q_3$.}  $Q_3$ 
are computed at next-to-leading (NLO) in $1/N_c$ within  the ENJL model.
The split in Penguin-like and $B_K$-like contributions is discussed.
For $Q_5$ we cannot simply discuss this split, here the correct
chiral behaviour is only reproduced after summing both contributions.

When extending the method to $Q_6$ one discovers that the factorizable
contribution from $Q_6$ has an infrared divergence in the chiral limit.
We discuss this problem in Section \ref{Q6discussion} and show how
it is cancelled by  the non-factorizable contribution.
This problem might be part of the reason why estimates for the $Q_6$ operator
vary so widely.
After correcting for this we present also results for the
matrix elements of $Q_6$.

Finally we put the numerical results for the long- and short-distances
together in Section \ref{fullresults} and discuss their stability.
We also discuss here the coefficients $a$, $b$, and $c$ defined earlier
by Pich and de Rafael \cite{PdeR95}. We recapitulate our main results
and conclusions in Section \ref{conclusions}

\section{The $\Delta I=1/2$ Rule in $K \to \pi \pi$}
\label{deltaI}

The $K\to\pi\pi$ invariant amplitudes can be decomposed into definite isospin 
quantum numbers amplitudes as $[A\equiv -i T]$
\ba
A[K_S\to \pi^0\pi^0] &
\equiv &\sqrt{2\over3} A_0 -{2\over\sqrt 3} A_2 \nonumber \, ; \\
A[K_S\to \pi^+ \pi^-] 
&\equiv &\sqrt{2\over3} A_0 +{1\over\sqrt 3} A_2 \nonumber \, ; \\
A[K^+\to\pi^+\pi^0] &\equiv& {\sqrt{3}\over2} A_2 \, .
\ea
Where $K_S \simeq K_1^0 +\epsilon \, K_2^0$, $K^0_{1(2)}\equiv(K^0-(+)
\overline{K^0})/\sqrt 2$, and CP($K^0_{1(2)})=+(-)K^0_{1(2)}$.
In this paper we are interested in 
the CP conserving part of $K\to \pi\pi$, so we set 
the small phase in the Standard Model CKM matrix elements 
and therefore $\epsilon$ to zero.
Above we have included the final state interaction phases 
$\delta_0$ and $\delta_2$ into the
amplitudes $A_0$ and $A_2$ as follows. For the isospin $1/2$
amplitude
\be
A_0\equiv -i a_0 \, e^{i\delta_0}\, ,
\ee
and for the isospin $3/2$
\be
A_2\equiv -i a_2 \, e^{i\delta_2}\, .
\ee

With the measured $K_S\to \pi^0\pi^0$ partial width $\Gamma_{00}$,
  $K_S\to \pi^+\pi^-$ partial width $\Gamma_{+-}$,
and $K^+\to\pi^+\pi^0$  partial width $\Gamma_{+0}$ \cite{PDG}, 
we can calculate the ratio
\be
\left|\frac{A_0}{A_2}\right| = \left({3\over4}
\sqrt{\frac{1-4 m_\pi^2/m_{K^+}^2}{1-4m_\pi^2/m_{K^0}^2}}
\left(\frac{\Gamma_{00}+\Gamma_{+-}}{\Gamma_{+0}}\right)-1\right)^{1/2} = 22.10
\ee
This result is what is called the $\Delta I=1/2$ rule for kaon decays.

To understand quantitatively  this rule has been one of the permanent 
issues in the literature since the experimental determination.
It is by now clear
that it is the sum of several large contributions 
both from short distance origin \cite{one-loop,two-loops}
and from long distance origin \cite{BG87,BBG,KMW90}
which add constructively to make $|A_0|$ much larger than $|A_2|$. 

The lattice QCD community has also spent a large effort
on this problem, see \cite{lattice} for some recent reviews.

Among the long distance
enhancements of the $|A_0/A_2|$ ratio, the order $p^4$ chiral corrections
have been found to be quite important. 
The CHPT analysis to order $p^4$ can be found in \cite{KMW90}
and both the counter-terms and the chiral logs to that order
can be found in \cite{BPP98}, the chiral logs
were originally calculated in \cite{Bloos}. There are some small differences
between the two results. The fit of the data to both
the order $p^4$ $K\to \pi\pi$ and $K\to\pi\pi\pi$ counter-terms
and chiral logs \cite{KMW90,KAM90} allowed to extract 
\footnote{The fit uncertainties to this result were not quoted
 in \cite{KMW90,KAM90}.}
\ba
\label{I=1/2p2}
\left|\frac{A_0}{A_2}\right|^{(2)}&=& 16.4
\ea
to $O(p^2)$, i.e., around  34 \% of the enhancement in the $\Delta I=1/2$
rule is due just to order $p^4$ and higher CHPT corrections.

\subsection{CHPT to order $p^2$}
To order $p^2$ in CHPT, the amplitudes 
$a_0$ and $a_2$ can be written in terms of  two couplings,
\ba
a_0 \equiv a_0^{8}+a_0^{27}&=&C \, \left[9 G_8+G_{27}\right]
\frac{\sqrt 6}{9} F_0 (m_K^2-m_\pi^2)\, , \nonumber \\
a_2&=& C \, G_{27} \frac{10\sqrt 3}{9} F_0 (m_K^2-m_\pi^2)\, , 
\ea
with
\be
\label{defC}
C\equiv-\frac{3}{5}\, \frac{G_F}{\sqrt 2} \, V_{ud} \, V_{us}^*
\approx  - 1.06\cdot10^{-6} \, \mbox{GeV}^{-2}
\ee
and 
\be
\delta_0=\delta_2=0\, .
\ee
The couplings $G_8$ and $G_{27}$ are two of the $O(p^2)$ 
$\Delta S=1$ couplings. They are defined in \cite{BPP98} and can
be determined from the $O(p^2)$ amplitudes \cite{KAM90} to be
\be
\label{valueG8G27}
G_8 = 6.2 \pm 0.7  \qquad {\rm and} \qquad G_{27} = 0.48 \pm 0.06 \,.
\ee
Here we have  only included the error bars from the value 
of the pion decay constant in the chiral limit $F_0=(86\pm10)$ MeV,
this corresponds to $f_\pi=92.4$ MeV.
Again there are uncertainties from the fit procedure and approximations 
not quoted in \cite{KMW90,KAM90}.

Therefore to  $O(p^2)$
\be
\label{ratioA0A2p2}
\left|\frac{A_0}{A_2}\right|^{(2)}=\sqrt 2\, 
\left( \frac{9 \, G_8 + G_{27}}{10 \, G_{27}} \right)\, . 
\ee
To understand the difficulty of the task of reproducing
(\ref{I=1/2p2}) it is convenient to make an $1/N_c$ analysis
of the $O(p^2)$ result. At large $N_c$, $G_8=G_{27}=1$ and 
\be
\label{I=1/2largeN}
\left|\frac{A_0}{A_2}\right|^{(2)}_{N_c} = \sqrt 2 
\ee
i.e. a factor 11.6 smaller than the QCD result in (\ref{I=1/2p2}) !
Notice that to $O(p^2)$ there are no quark mass
and therefore no chiral logs corrections to the ratio above. 
So we have to explain one order of magnitude enhancement within QCD in the 
chiral limit with $1/N_c$ suppressed corrections.  

\begin{figure}
\begin{center}
\epsfig{file=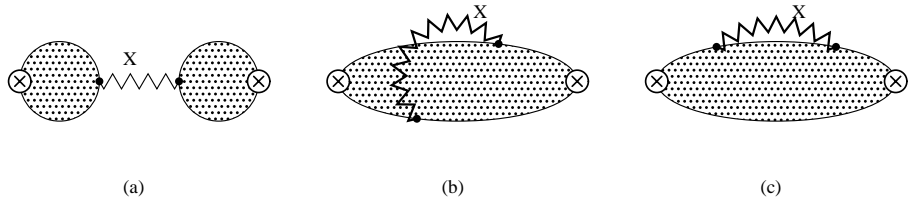,width=12cm}
\end{center}
\caption{\label{figfull} The three types of contributions appearing
in the evaluation of matrix-elements of operators. Namely, (a) Factorizable,
(b) $B_K$-like, and (c) Penguin-like.}
\end{figure}

Another parametrization which will be useful when
studying the $\Delta I=1/2$ rule is the one introduced
by Pich and de Rafael in \cite{PdeR95}. In this parametrization
\ba
\label{abc}
G_{27}&\equiv& a + b \, , \nonumber \\
G_{8}&\equiv& a + b + {5\over3}(c-b) \, .
\ea
The nice feature of this parametrization is that $a$, $b$, and
$c$ have a one to one correspondence with the three-different
QCD quark-level topologies. The $a$-type coupling  corresponds to 
configurations that include the factorizable ones (Figure \ref{figfull}a).
 This coupling is of
order 1 in the large $N_c$ limit and has only $1/N_c^2$
corrections. The $b$-type coupling corresponds to what we call $B_K$-like
topologies (Figure \ref{figfull}b) and is of order $1/N_c$.
This coupling is related to the value of the $B_K$ parameter 
in the chiral limit. 

The $c$-type coupling corresponds to what we call Penguin-like topologies
(Figure \ref{figfull}c) and is also of order $1/N_c$.
So in the large $N_c$ limit
\be
a=1 \hspace{1cm} {\rm and} \hspace{1cm}\; b=c=0 \; .
\ee
The main objective of this paper is the calculation of  the $1/N_c$ 
corrections to (\ref{I=1/2largeN}), i.e. the couplings $b$ and $c$.

The coefficients $a$, $b$, and $c$ in \cite{PdeR95} were defined in a
 large $N_c$ expansion within short-distance QCD, i.e.
with quarks and gluons. In the low-energy regime where
the long-distance part has to be evaluated
 one  however cannot distinguish the 
$1/N_c^2$ corrections to $a$ from the ones to the coefficients $b$
and $c$. So for us $a$ takes the large $N_c$ value
$a=1$, 
$b=G_{27}-1$, and $c=(3G_8+2G_{27})/5-1$. This definition can be used
both at long and short-distances and only differs by terms
of $O(1/N_c^2)$ with the one in \cite{PdeR95}.
The definition above has also the advantage that all couplings $a$,
$b$, and $c$ are scale independent. 
Notice that in the present work  the $1/N_c^2$
are from short-distance origin only.

\section{The Technique}

\subsection{$\Delta S=1$  and $\Delta S=2$ Two-Point Functions}
\label{twopdef}

The theoretical framework we use to study
the  strangeness changing transitions in one and two units
was already introduced in Refs. \cite{BPBK,BPP98,BPDashen}.
The original suggestion for this type of method was in \cite{BDSPW85}.
The basic objects are the pseudo-scalar density correlators
\be
\label{two-point}
\Pi^{ij}(q^2)\equiv 
i \int {\rm d}^4 x \, e^{i q.x} \langle 0 |
T\left(P^{i\dagger}(0) P^j(x) e^{i \Gamma_{\Delta S=a}}\right) |0\rangle
\ee
in the presence of strong interactions. Above, $a=0, 1, 2$
stands for $|\Delta S| =$ 0, 1, and 2 transitions and $i, j$ are light quark
combinations corresponding to the octet of the lightest
pseudo-scalar mesons;
\ba
\label{pseudosources}
P^{\pi^0}(x)\equiv \frac{\dis 1}{\dis \sqrt 2}  
\left[\overline u i \gamma_5 u
- \overline d i \gamma_5 d\right]\, ,  &
P^{\pi^+}(x)\equiv \left[ \overline d i \gamma_5 u \right]\, ,  
P^{K^0}(x)\equiv \left[ \overline s i \gamma_5 d \right] \, , \nonumber \\
P^{K^+}(x)\equiv \left[\overline s i \gamma_5 u \right]\, , & 
P^{\eta_8}(x)\equiv  \frac{\dis 1}{\dis \sqrt 6}  
\left[\overline u i \gamma_5 u
+ \overline d i \gamma_5 d - 2 \overline s i \gamma_5 s \right]\, .
\nonumber \\  
\ea
Here and in the remainder, summation over colour-indices inside brackets
is assumed unless colour indices have been explicitly indicated.
These two-point functions
were analyzed extensively within CHPT to order $p^4$ in \cite{BPP98}.
In that reference we also pointed out how on can obtain information
on $K \to \pi \pi$ amplitudes 
at order $p^4$ from  off-shell $K \to \pi$ transitions.

Now, we want to use the $1/N_c$ technique used in \cite{BPBK,BPDashen}
to compute the off-shell $K \to \pi$ amplitudes and obtain
the relevant counter-terms of order $p^2$.
See  \cite{BPP98}, 
for explicit details of which counter-terms of order $p^4$ we can get
and possible ways of estimating some couplings  we cannot
get this way.

In the large $N_c$ limit, there is just one operator
in the Standard Model which changes strangeness by one-unit 
\be
\label{q2}
Q_2 \equiv [\bar{s}\gamma^\mu(1-\gamma_5)u](x)
[\bar{u}\gamma_\mu(1-\gamma_5) d](x) \, .
\ee
After the inclusion of gluonic corrections $Q_2$ mixes with 
\be
\label{q1}
Q_1 \equiv [\bar{s}\gamma^\mu(1-\gamma_5)d](x)
[\bar{u}\gamma_\mu(1-\gamma_5) u](x)
\ee
via box-type diagrams (first reference in \cite{one-loop}), 
and with 
\ba
\label{operators}
Q_3&\equiv& [\bar{s}\gamma^\mu(1-\gamma_5)d](x)
\sum_{q=u,d,s}[\bar{q}\gamma_\mu(1-\gamma_5) q](x)
\nonumber\\ \nonumber
Q_4&\equiv& [\bar{s}^\alpha\gamma^\mu(1-\gamma_5)d_\beta](x)
\sum_{q=u,d,s}[\bar{q}^\beta\gamma_\mu(1-\gamma_5) q_\alpha](x)
\\ \nonumber
Q_5&\equiv& [\bar{s}\gamma^\mu(1-\gamma_5)d](x)
\sum_{q=u,d,s}[\bar{q}\gamma_\mu (1+\gamma_5)q](x)
\\ 
 Q_6&\equiv& [\bar{s}^\alpha\gamma^\mu(1-\gamma_5)d_\beta](x)
\sum_{q=u,d,s}[\bar{q}^\beta\gamma_\mu(1+\gamma_5) q_\alpha](x)
\ea
via the so-called penguin-type diagrams \cite{one-loop}. Since the numerical
importance for the issues we want to address here 
 is small and for the sake of
simplicity we  switch off electromagnetic  interactions. The operator
$Q_4$ is redundant and satisfies $Q_4 = Q_2-Q_1+Q_3$. 
Under SU(3)$_L$$\times$SU(3)$_R$ rotations $Q_-
\equiv Q_2-Q_1$, $Q_3$, $Q_4$, $Q_5$,
and $Q_6$ transform as $8_L \times 1_R$ and only carry
$\Delta I=1/2$ while $Q_{27}\equiv 
3 Q_1 + 2 Q_2 - Q_3$ transforms as $27_L \times 1_R$ and carries
both $\Delta I=1/2$ and $\Delta I=3/2$. 

The Standard Model low energy effective action describing
 $\left|\Delta S\right|=1$  transitions can thus be written as
\be
\Gamma_{\Delta S=1} \equiv -C_{\Delta S=1} 
\, {\dis \sum_{i=1}^6} \, C_i(\mu) \, 
\int {\rm d}^4 y  \, Q_i (y) \, + {\rm h.c.}
\ee
where
$C_{\Delta S=1} = (G_F/\sqrt 2) \, V_{ud} V_{us}^* \,$.

There is just one operator changing strangeness by two-units
in the Standard Model, 
\be
\label{qS2}
Q_{\Delta S=2}\equiv [\bar{s}\gamma^\mu(1-\gamma_5)d](x)
[\bar{s}\gamma_\mu(1-\gamma_5)d](x)
\ee
 which transforms under SU(3)$_L$$\times$SU(3)$_R$ rotations 
as $27_L \times 1_R$.

The matrix elements of the $Q_i$ with $i=1,\cdots,6$, and $Q_{\Delta S=2}$
operators depend on the renormalization
group (RG) scale $\mu$ such that physical processes are scale independent. 

\subsection{The $X$-Boson Method and Matching}
\label{Xboson}
In this section we explain the basics of  how to deal with the
resummation of
large logarithms using the renormalization group and how to
do the matching between the low energy model and the short-distance evolution
inside QCD. The guiding line here is the $1/N_c$ expansion.

Let us first explain the philosophy in the case of photon non-leptonic
processes \cite{BPDashen,BBGpp,BDashen}. The basic electromagnetic (EM)
non-leptonic interaction is given by
\be
{\cal L}_{EM} =
\frac{(ie)^2}{2} \, \int \frac{{\rm d}^4 r}{(2\pi)^4}
\int {\rm d}^4 x \, \int {\rm d}^4 y \,  e^{i q\cdot (x-y)} 
\frac{g_{\mu\nu}}{r^2-i\epsilon} J^\mu_{Had}(x)\, J^{\nu}_{Had}(y)\,.
\ee
Here we used the Feynman gauge, for a discussion of the gauge dependence
see \cite{BPDashen}, $J^\mu=(\overline q Q \gamma^\mu q)$,
$q^T = (u, d, s)$ and $Q$ is a 
3 $\times$ 3 diagonal matrix collecting the light quark electric charges.
The integral over $r^2$ we rotate into Euclidean space and split into
a long and a short distance piece,
\be
\label{split}
\int {\rm d}^4 r_E = \int {\rm d} \Omega
\left(\int_0^{\mu} {\rm d} |r_E| \, |r_E|^3 + 
\int_{\mu}^\infty {\rm d} |r_E| \, |r_E|^3 \right)\,.
\ee
The long distance piece we evaluate
in an appropriate low-energy model, CHPT\cite{BDashen}, ENJL\cite{BPDashen}
or using other hadronic models \cite{BBGpp}.
The short-distance part can be evaluated using the operator product expansion
(OPE) and the matrix-elements of the resulting operators can be evaluated to 
the leading non-trivial order in $1/N_c$ using the same hadronic 
low-energy hadronic model as for the long-distance part.

This procedure works extremely well in the case of internal photon exchange.
The problem is that in weak decays there are large logarithms
present of the type $\ln(M_W/\mu_L)/N_c$ which make the
$1/N_c$ expansion of questionable validity. The solution to this problem
at one-loop order was presented in \cite{BPBK} where we showed that
the integral in (\ref{split}) satisfied the same equation as the one-loop
evolution equation. This method was very nice for $B_K$ and can
also be applied to the $\Delta S=1$ transitions.

Here we will give an alternative description of the method used there
that will be extendable in a relatively straightforward way to the
two-loop renormalization group calculations. The precise definition
and calculations we defer to a future calculation.

We start at the scale $M_W$ where we replace the exchange of $W$ and top
quark in the full theory with higher dimensional
operators using the OPE in  an effective theory 
where these heavy particles have  been integrated out. 
So at a scale $\mu_H\approx M_W$ we need the matching conditions between
the full theory and the effective one. As usual we get them  by 
setting the matrix elements between external states of light particles,
i.e. the remaining quarks and gluons, in transition amplitudes with
 $W$ boson and top quark exchanges 
equal to those of the relevant operators in the effective theory.
\be
\mbox{Step 1: at }\mu_H\approx M_W~:\quad \\ 
\langle 2 | (W,top\mathrm{-exchange})_{Full}|1\rangle = 
\langle 2|\sum_i \, \tilde C_i(\mu_H) \, \tilde Q_i |1\rangle \, . 
\ee

We then proceed by using the renormalization group to run down from $\mu_H$
to $\mu_L$ below the charm quark mass where we have an effective theory
with gluons and the three lightest quark  flavours.
At each heavy particle threshold crossed new matching conditions 
between the two effective field theories (with and without the heavy particles
being integrated out) have to be  set,  this is done completely within 
perturbative QCD, see e.g. \cite{BurasReviews}. So that
\be
\mbox{Step 2: from }\mu_H ~\mathrm{to~}\mu_L \qquad  
\langle 2|\sum_i \, \tilde C_i(\mu_H) \, \tilde Q_i |1\rangle  
\longrightarrow
\langle 2|\sum_j \, C_j(\mu_L) \, Q_j |1\rangle \, . 
\ee

At Step 3 we again introduce a new effective field theory which reproduces
the physics of the  operators $Q_j$ below $\mu_L$ by  the exchange
of heavy $X_i$-bosons with  couplings $g_i$. Again we need to set 
matching conditions
\be
\label{match3}
\mbox{Step 3: at }\mu_L :\quad 
\langle 2 | (X_j\mathrm{-exchange})|1\rangle = 
\langle 2|\sum_j C_j(\mu_L) Q_j |1\rangle\,.
\ee
Here the matching means that the left hand side should be evaluated in an
operator product expansion in $M_{X_i}$

The right hand side matrix elements in (\ref{match3}) can be evaluated
completely within perturbative QCD and therefore all
the dependence on the renormalization scheme and the choice
of the basis $Q_j$ and of evanescent operators  disappears in this step.
This procedures fixes the $g_i$ couplings as functions of the
chosen masses $M_{X_i}$ and the matrix elements 
$\langle 2|\sum_j C_j(\mu_L) Q_j |1\rangle$ which are scheme
independent. Depending on the order to which we decide to calculate in the 
effective theory, $g_i$ will depend on  additional
terms that can be fully determined within the effective 
theory with heavy $X_i$ bosons. 

\begin{figure}
\begin{center}
\epsfig{file=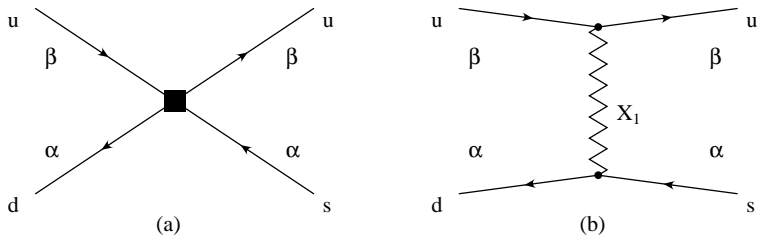, width=10cm}
\end{center}
\caption{\label{figX} The reproduction of the operator $Q_1$
by the exchange of a neutral boson $X_1$.}
\end{figure}

As an example, let us use the effective field theory with  two-loop 
accuracy for  the running between
scales $\mu_H$ and  $\mu_L$ and calculations at  next-to-leading order 
in $1/N_c$  within the heavy $X_i$ boson effective theory. 
The term $C_1(\mu_L) \, Q_1$ is reproduced  in the $X_i$ 
effective field theory by the exchange of a  heavy enough 
vector-boson $X_1^\mu$ with couplings 
\be
X_1^\mu \left\{ g_1 \left[\bar{s} \gamma_\mu (1-\gamma_5) d \right]+
g_1^\prime \left[\bar{u} \gamma_\mu (1-\gamma_5) u \right]
\right\}\, + {\rm h.c.}. 
\ee 
The $X_1$ boson has only $\Delta S=1$ components.
This is shown pictorially
in Fig. \ref{figX}.
The scale $\mu_L$ should be high enough to use 
perturbation theory.
We have the following  matching conditions (\ref{match3}) 
in this case (we assume that $Q_1$ only has 
multiplicative renormalization for simplicity)
\be
\label{match4}
\frac{g_1  \,g_1'^\dagger }{M_{X_1}^2} \left(
1+\frac{\alpha_s(\mu_L)}{\pi}\left[\tilde d_1
\ln\left(\frac{M_{X_1}}{\mu_L}
\right)+\tilde r_1\right]\right)=
C_1(\mu_L) \left( 1 + \frac{\alpha_s(\mu_L)}{\pi}\,  r_1 \right) \, .
\ee
The  $r_1$ term cancels the scheme dependence of the two-loop
Wilson coefficient $C_1(\mu_L)$.
Notice that we can choose independently 
any regularization scheme on the left and right hand sides.
In the present work we will use the NDR (naive dimensional regularization)
two-loop running between $\mu_W$ and $\mu_L$.
All the large logarithms of the type
$\ln(M_W/\mu_L)$ are absorbed in the couplings of the $X_i$
boson in a scheme independent way.

Now we come to Step 4.
Assume we want to calculate $K^0 \to \pi^0$ matrix element
in the Standard Model. Since we have included the effect of all the large 
logarithms between $M_W$ and $\mu_L$ in the $g_i$ couplings, 
we can now apply the same procedure explained at the beginning of this section
for the  photon exchange case \cite{BPDashen,BBGpp,BDashen}
and remain at next-to-leading order in $1/N_c$. This we do
now for the  effective three-flavour 
field theory with heavy massive $X_i$ bosons. 
So we split the integral over $|r_E|$ into a long distance
piece (between 0 and $\mu$) and a short distance piece (between $\mu$ and
$\infty$) as in (\ref{split}).
When evaluating the second term in (\ref{split}) we will find precisely
the correct logarithmic dependence on $M_{X_1}$ to cancel the one in
(\ref{match4}). The presentation of the scheme dependent
constants $r_1$ and $\tilde r_1$ for $\Delta S=1$ and $\Delta S=2$
is deferred to a future publication.

We then require some matching window in $\mu$ 
along the lines explained in \cite{BPBK} between these two pieces.
We will use the framework described above
to calculate $\Delta S=1$ and $\Delta S=2$ two-point functions
and defer the full discussion about this procedure to a future publication.
In practice we will also choose $\mu=\mu_L$.

The same procedure can in principle be used in lattice gauge theory
calculations where one can then include the $X_i$-bosons explicitly in the
lattice regularized theory or equivalently work with the corresponding
non-local operators.

\subsection{The Low-Energy Model}
\label{ENJL}
The low-energy model we use here is the extended Nambu--Jona-Lasinio model.
It consists out of the free lagrangian for the quarks
with point-like four-quark couplings added. This model has the correct
chiral structure and spontaneously breaks
chiral symmetry. It includes a surprisingly large amount of the observed
low energy hadronic phenomenology. We refer to the review articles
\cite{reviewsNJL} and the previous papers where we have discussed
the various aspects of the ENJL model used
here \cite{BPBK,BBR,BPano,BP94,BPPgm2}. A short overview of
the advantages and disadvantages can be found in \cite{BPDashen}
Section 3.2.1.

It is well known however that it doesn't confine and 
doesn't have the correct momenta dependence at large $N_c$ in some cases.
These two issues were treated in  \cite{PPR98} were a low energy model
correcting  the wrong momenta dependence at  large $N_c$ was presented.

The bad high energy behaviour of ENJL two-point functions
produces some unphysical cut-off dependence. 
In this work we try to smear out this bad behaviour as follows.
For the fitting procedure 
we only use  points with small values of all momenta and
always Euclidean. We also keep only the few first terms
in the fit to a polynomial ( of order six at most)
which are therefore not extremely sensitive
to the bad high energy behaviour of the ENJL model. 
The model in \cite{PPR98} gives very good
perspectives that this unphysical behaviour can be eliminated to a  large
extent,  see for instance the
recent work in \cite{KPR98},  and  would  provide a natural
extension of this work. 

\section{$\Delta S=2$ Transitions: Long Distance}
\label{BKsect}

In this section we apply the technique to
$\Delta S=2$ transitions. These transitions were already studied
in \cite{BPBK} using the same model for the low
energy contributions, there are however differences in the routing
of the momenta with respect to the one we took in \cite{BPBK}.
See the next section for a discussion of this issue.

We study the two-point function
$\Pi^{\overline K^0 \, K^0}(q^2)$ in the presence of strong interactions
as defined in (\ref{two-point}). 
The operators in $\Gamma_{\Delta S =2}$ are replaced by an $X$ boson
coupling to $[\bar{s}\gamma_\mu(1-\gamma_5)d](x)$ currents as described in
Section \ref{Xboson}.

We evaluate the two-point function then as a function of $\mu$ for various
values of $q^2$ and masses and this allows us to extract the relevant
couplings in CHPT.  We restrict ourselves here to the $O(p^2)$ coefficient
$G_{27}$ and the actual value of $\hat B_K$.

\subsection{The Routing Issue}
\label{routing}
In this section we would like to explain why our present results
on $B_K$ differ from those presented in \cite{BPBK} even though we
use the same method and the same model. At the same time this
will explain the difference between the result from 
 Section  4 in \cite{BPBK} for $G_{27}$ and the 
one  from \cite{FG95}. Both papers use the method of \cite{BBGpp}
and \cite{BGK91} to identify the cut-off scale used to identify with the
short-distance evolution and we have several times checked the calculations
in both papers and found no errors in either. We will present the discussion
here in the case where the low energy model used is CHPT to simplify
the discussion.

\begin{figure}
\begin{center}
\epsfig{file=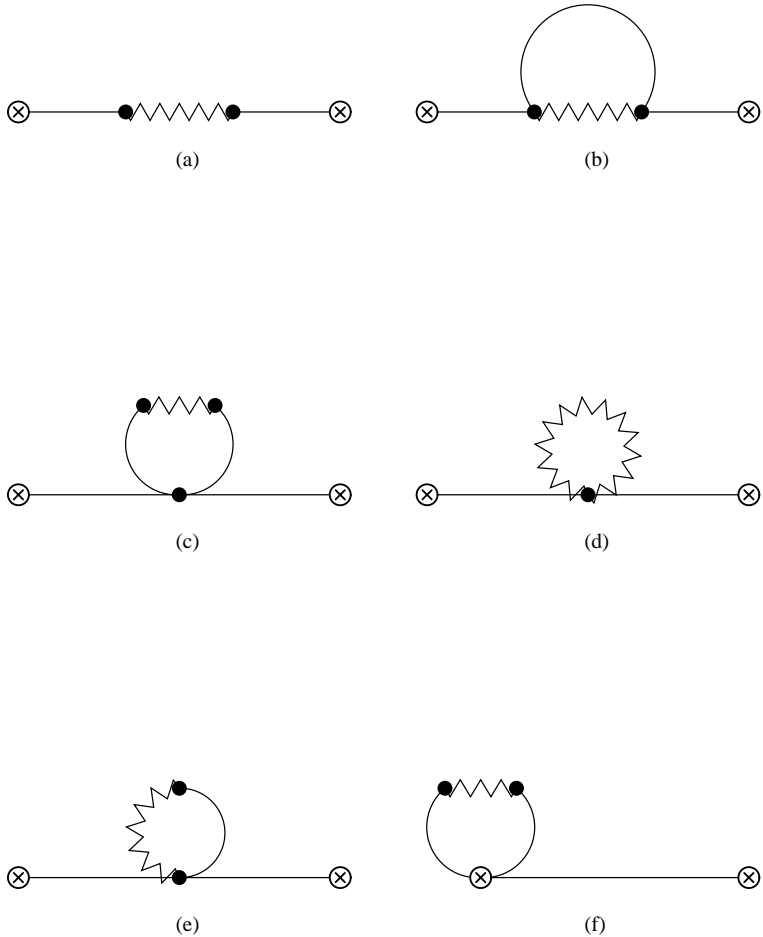,width=10cm}
\end{center}
\caption{\label{figBK} Chiral Perturbation Theory
 contributions to $\Pi_{\Delta S=a}(q^2)$.
(a) Lowest order. (b)-(f) Higher order non-factorizable.
The full lines are mesons. The zig--zag line is the $X$-boson.}
\end{figure}

\begin{figure}
\begin{center}
\epsfig{file={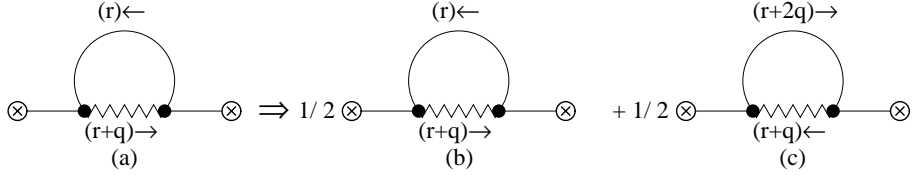},width=12cm}
\end{center}
\caption{\label{figrouting} The routing for the $\Delta S=2$ operator enforced
by CPS symmetry. (a) Routing used  in \cite{BPBK}, (b)+(c) 
The correct routing as it should have been used.}
\end{figure}

The source of the difference turned out to be more subtle. In \cite{BPBK}
the choice of momentum for the $X$-boson was made to be $r+q$ where
$q$ is the momentum going through the two-point function defined
in (\ref{two-point}) and $r$ is the loop integration variable.
This particular choice was
done in order to have the lowest order always non-zero, even if the
range of momenta in $r$ integrated over was such that $|r^2|<|q^2|$.
We had also always chosen the direction of $r+q$ through the $X$ boson such
that the internal propagator appearing in diagram (b) of Fig. \ref{figBK}
had momentum $r$. Since the $X$ in that case was a neutral gauge boson this
was a natural choice. 
It turns out however that in the presence of a
cut-off some of the contributions obtained 
with this routing do not have the correct
CPS symmetry. This symmetry imposes that some of the contributions have to 
have the internal propagator in Fig. \ref{figBK}
with momentum $r+2q$ instead of $r$. The precise change has been depicted in
Fig. \ref{figrouting}. The momentum flow as depicted in (a) should be replaced
by the sum of (b) and (c).
This doesn't affect the coefficients of the chiral logarithms.
Therefore one can
use any  routing when using regularization which doesn't see analytic 
dependence on the cut-off. Unfortunately, this bad routing
was actually causing  most 
of the bad behaviour for $B_K(\mu)$ for high values of $\mu$
in Table 1 of \cite{BPBK} and the difference
with the result for $G_{27}$ of \cite{BPBK} and \cite{FG95}.
In fact, using the background
field method as in \cite{FG95} the CPS symmetry is automatically satisfied
at order $p^2$ with any routing. 

We have now corrected for this problem and obtain  a much more reasonable
matching  between long-distance contributions
and the short-distance contributions.
Nevertheless, it turns out that the range of values chosen for $\mu$ 
in \cite{BPBK} to make the predictions was not very much affected by
 the routing problem explained above. The results we now obtain
are much more stable numerically and  in the same ranges as the ones
quoted in \cite{BPBK}. We also agree with the result in \cite{FG95}
for $G_{27}(\mu)$ obtained from lowest order CHPT,
\be
\label{G27CHPT}
G_{27}(\mu) = 1 - \frac{3\mu^2}{16\pi^2 F_0^2}\,.
\ee
Here and in what follows,  the $\mu$ dependent $G_8(\mu)$, $G_8'(\mu)$,
$G_{27}(\mu)$, and $B_K(\mu)$ couplings stand for the long-distance
contributions to those couplings, i.e. with 
$[1+(\alpha_s(\mu)/\pi)\,  r_{1, j}] \, C_j(\mu)=1$.

\subsection{CHPT Results}
Here we update Section 4 of \cite{BPBK} to correct for the routing
problem. The non-factorizable contribution to
$\Pi^{\overline K^0 \, K^0}(q^2)$ is given by the diagrams in Figure
\ref{figBK} and is:
\ba
\lefteqn{\frac{-8 B_0^2 F_0^2}{(q^2-m_K^2)^2} \Bigg\{
\int^\mu\frac{d^4r_E}{(2\pi)^4}\frac{r_E^2 q_E^2}{(r_E^2+m_K^2)^2}
-\int^\mu\frac{d^4r_E}{(2\pi)^4}\frac{r_E^2}{r_E^2+m_K^2}}&&\nonumber\\&&
\!\!\!+\frac{1}{2}\int^\mu\frac{d^4r_E}{(2\pi)^4} (r_E+2 q_E)^2
\left[\frac{1}{(r_E+q_E)^2+m_\pi^2}+\frac{1}{(r_E+q_E)^2+2m_K^2-m_\pi^2}\right]
\Bigg\} \nonumber \\
\ea
These integrals can be performed analytically but the result is rather
cumbersome.  The Euclidean continuation of $q^2$ we used is $q_E^2 = -q^2$. 
The result in the chiral limit becomes
\be
\frac{-8 B_0^2 F_0^2}{(q^2-m_K^2)^2}\frac{1}{16\pi^2F_0^2}
\left\{-3\mu^2 q^2-\frac{5}{6}q^4\right\}
\ee
and for $q^2$ = 0
\ba
\frac{-8 B_0^2 F_0^2}{(q^2-m_K^2)^2}\frac{1}{16\pi^2F_0^2}
  \nonumber &&\\ && \hspace{-4cm} \times \Bigg\{
-\frac{1}{2}(2m_K^2-m_\pi^2)\left(\mu^2-(2m_K^2-m_\pi^2)
 \ln\left(\frac{\mu^2+2m_K^2-m_\pi^2}{2m_K^2-m_\pi^2}\right)\right) 
\nonumber \\ && \hspace{-4cm} +
m_K^2\left(\mu^2-m_K^2\ln\left(\frac{\mu^2+m_K^2}{m_K^2}\right)\right)
-\frac{1}{2}m_\pi^2
\left(\mu^2-m_\pi^2\ln\left(\frac{\mu^2+m_\pi^2}{m_\pi^2}\right)\right)
\Bigg\} \nonumber \\
\ea
These results allow to obtain the equivalent of (\ref{G27CHPT})
for the $O(p^4)$ coefficients.

\subsection{The $B_K$ Parameter: Long Distance and Short Distance}
We now take the results from the ENJL evaluation of
$\Pi^{\overline K^0 \, K^0}(q^2)$ both in the chiral limit and
in the case of quark masses corresponding to the physical pion and
kaon mass and use these to estimate $B_K$ and $G_{27}$.

The final results for $B_K$ in the chiral limit, $B_K^\chi(\mu)$
and $G_{27}(\mu)=4 B_K^\chi(\mu)/3$ are shown in Table \ref{tableBK}.
\begin{table}
\begin{center}
\begin{tabular}{|c|ccccccc|}
\hline
$\mu$(GeV)&$G_{27}(\mu)$&$B_K^\chi(\mu)$&$B_K(\mu)$&$\hat B_{K(1)}$
&$\hat B_{K(2)}^{\mbox{SI}}$&$\hat B_{K(2)}^{\exp}$&$
\hat B_{K(2)}^{\chi\exp}$\\
\hline
0.3& 0.830 & 0.622 & 0.784 & --   & --   & --   & --\\
0.4& 0.737 & 0.552 & 0.776 & --   & --   & --   & --\\
0.5& 0.638 & 0.478 & 0.762 & 0.79 & 0.36 & 0.48 & 0.30\\
0.6& 0.537 & 0.402 & 0.746 & 0.81 & 0.57 & 0.62 & 0.33\\
0.7& 0.431 & 0.323 & 0.721 & 0.81 & 0.63 & 0.66 & 0.30\\
0.8& 0.320 & 0.240 & 0.688 & 0.79 & 0.65 & 0.67 & 0.23\\
0.9& 0.200 & 0.150 & 0.643 & 0.75 & 0.64 & 0.66 & 0.15\\
1.0& 0.070 & 0.052 & 0.588 & 0.70 & 0.61 & 0.62 & 0.05\\
\hline
\end{tabular}
\end{center}
\caption{\label{tableBK} The long-distance contributions 
to $G_{27}(\mu)$, $B_K^\chi(\mu)$ and $B_K(\mu)$
as determined using the ENJL model. Also shown are 
$\hat B_{K(1)}$ using the one-loop short distance
and $\hat B_{K(2)}$, $\hat B_{K(2)}^{\chi}$ using the  
two-loop short distance in Table \ref{WilsonS=2}.
See  Appendix \ref{AppA} for the values of the
parameters used. For the non-chiral cases one has to
add 0.09$\pm$0.03 from the nonet vs octet difference, see text.}
\end{table}
We have also shown the value of $B_K$ obtained there from extrapolating the
ENJL two-point function in the Euclidean domain to the kaon pole
using Chiral Perturbation Theory, this is $B_K(\mu)$.
In the latter case we have to include the correction
due to the difference between the octet and nonet case. This
correction was estimated to be about $0.09\pm0.03$ in \cite{BPBK}
and we take it as $\mu$-independent.
In the other columns in Table \ref{tableBK} various parts of
the short distance correction are included. The realistic case, with non-zero
quark masses in the long distance contribution to
$B_K(\mu)$, we have shown with
the one-loop short-distance running, $\hat B_{K(1)}$, two-loop short
distance running with the scheme-dependence removed, $\hat B^{SI}_{K(2)}$,
as defined in eq. (\ref{defBKhat}), and the exact solution to the
two-loop evolution equation with the scheme-dependence removed
to the same order, $\hat B^{exp}_{K(2)}$ as defined
in Eq. (\ref{defBKhatexp}). 
For the latter short-distance contribution we have also shown the
result in the chiral limit, $\hat B_{K(2)}^{\chi exp}$. The rest
of the parameters used are in App. \ref{AppA}.
Notice that the matching for all cases is acceptable. The quality
of the matching for the real $\hat B_K$ is as good as for $\hat B^{exp}_{K(2)}$
since they only differ by the $\mu$-independent correction of 0.09 described
above.

So, in the chiral limit we get
\be
0.25<  \hat B_K^\chi < 0.40 \, , 
\ee
with non-zero quark masses we get
\be
0.50 < \hat B_K^{\rm Nonet} < 0.70
\ee
for the nonet case  and 
\be
0.59<  \hat B_K < 0.79  
\ee
for the real case.  
Notice that the large value of the chiral symmetry breaking
ratio
\be
 1.8 <  \frac{B_K}{B_K^\chi} < 2.4 \,  
\ee 
confirms the qualitative picture obtained in  \cite{BPBK}.
Finally, let us split the different contributions to the 
value of $\hat B_K$ in the real case, 
\be
\label{splitBK}
\hat B_K = (0.33\pm0.10) + (0.09\pm0.02) + 
(0.18\pm0.07) + (0.09\pm0.03)
\ee
where the first terms is the chiral limit result, the second term
are the $O(p^4)$ chiral logs at $\nu=M_\rho$ \cite{BPBK}, the
third term are the $O(p^4)$ counterterms and higher also
at the same scale 
and the last term is the above mentioned contribution
due to the $\eta_1-\eta_8$ mixing \cite{BPBK}. The error on the
chiral log contribution is from varying $\nu$ and the one on
the counterterm contribution from looking at various different
ways to extract the same counterterms as described in \cite{BPBK}
plus some extra as an estimate of the model error.

Notice that the last term in (\ref{splitBK}) is of order $(m_s-m_d)/N_c^2$
 and it is not included in present lattice results. In fact, it introduces
an unknown systematic uncertainty in quenched and partially unquenched 
results which is difficult to pin down, see \cite{lattice}.
So the lattice results cannot be easily compared to ours.
This term is also not included in determinations which are based in
lowest order CHPT since it is higher order. Therefore again a direct
comparison with  our results has to be done carefully. 

\section{$\Delta S=1$ Transitions: Long Distance}
\label{deltaS1longdistance}

In this section we use the $\Delta S=1$ two-point-functions
$\Pi^{K^+\pi^+}(q^2)$ and $\Pi^{K^0\pi^0}(q^2)$ as defined in
(\ref{two-point}). We do not use the one with $\eta_8$ since to
order $p^2$ we do not get any more information out of that
two-point function. It will provide extra information  to  $O(p^4)$
\cite{BPP98}.
 The result to lowest order in CHPT is given by
\ba
\!\!\Pi^{K^+\pi^+}(q^2)&=&-\frac{4 B_0^2 F_0^4\,C}
{(q^2-m_K^2)(q^2-m_\pi^2)}
\left[q^2\left(G_8+\frac{2}{3}G_{27}-2 G_8^\prime\right)
+m_\pi^2 G_8^\prime\right]\nonumber\\
\hskip-0.3mm\Pi^{K^0\pi^0}(q^2)&=&-\frac{2 \sqrt{2} B_0^2 F_0^4\,C }
{(q^2-m_K^2)(q^2-m_\pi^2)}
\left[q^2\left(-G_8+G_{27}+2 G_8^\prime\right)
-m_\pi^2 G_8^\prime\right]
\ea
Here $C$ of Eq. (\ref{defC})
has been chosen such that in the strict large-$N_c$ limit 
$G_8 = G_{27} = 1$. The coupling 
$G_8^\prime$ is the coefficient of the weak mass
term that does not contribute to $K\to\pi\pi$ at order $p^2$ but its
value is important at $O(p^4)$ and higher and for some
processes involving photons.
The definition of the $O(p^2)$ Lagrangian, a discussion of the 
contributions from $G_8^\prime$ and further references 
can be found in \cite{BPP98}.

We have calculated the two-point functions in the chiral limit to extract
the coefficient of $q^2$ and in the
case of equal quark masses for an ENJL quark mass of
0.5, 1, 5, 10, and 20
MeV in order to extract the coefficient of $m_\pi^2$.

As described in Section \ref{Xboson} we treat all coefficients $C_i(\mu_L)$
as leading order in $1/N_c$ since they are enhanced in principle by large
logarithms. We therefore obtain the matrix elements of $Q_1$, $Q_2$,
$Q_3$, $Q_4=Q_2-Q_1+Q_3$, $Q_5$ and $Q_6$ to next-to-leading order
in $1/N_c$.

\subsection{Current x Current Operators}
The comments here are only valid for $Q_1, Q_2, Q_3, Q_4$, and
$Q_5$. The operator $Q_6$ is special
and is treated separately in the next subsection.

We can now use the method of \cite{BBG} with the correct routing and obtain
for the contributions to $G_8$ and $G_8^\prime$ from $Q_1$, $Q_2$,
$Q_3$ and $Q_5$ [with $\Delta_\mu\equiv \mu^2 /(16\pi^2F_0^2)$]:
\ba
\label{G8CHPT}
G_{27}(\mu)[Q_1]& = & 
G_{27}(\mu)[Q_2] = 1-3 \Delta_\mu + O(p^4) \, , \nonumber\\
G_8(\mu)[Q_1] & = & -\frac{2}{3}\left[1+\frac{9}{2} \Delta_\mu
 + O(p^4) \right]\, , \nonumber\\
G_8(\mu)[Q_2] & = & 1+\frac{9}{2}\, \Delta_\mu + O(p^4) \,, \nonumber\\
G_8(\mu)[Q_3] & = & 2 G_8(\mu)[Q_2] + 3 G_8(\mu)[Q_1]= 0+O(p^4)\,,\nonumber \\
G_8(\mu)[Q_5] &=& 0+O(p^4) \, , \nonumber\\
G_8^\prime(\mu)[Q_1] &=& 0 \, , \nonumber\\
G_8^\prime(\mu)[Q_2] &=& \frac{5}{6}\, \Delta_\mu + O(p^4) \, ,  \nonumber\\
G_8^\prime(\mu)[Q_3] & = & 2 G_8^\prime(\mu)[Q_2]=  
\frac{5}{3}\, \Delta_\mu + O(p^4)\, , \nonumber \\
G_8^\prime(\mu)[Q_5] &=& -\frac{5}{3} \Delta_\mu + O(p^4)\, . 
\ea
Here and in the remainder $G_8(\mu)[Q_i]$ stands for the long-distance
contribution of operator $Q_i$ to $G_8$ when $C_i(\mu)$ is set equal to 1.
The same definition applies to $G_8^\prime(\mu)[Q_i]$ and $G_{27}(\mu)[Q_i]$.
In Tables \ref{tablecurrent} and \ref{tableQ6} we dropped the
argument $(\mu)$ for brevity. 
The results from the ENJL calculations are summarized in Table
\ref{tablecurrent}.
The numbers in the columns 2 to 8 are always assuming
$[1+(\alpha_s (\mu)/\pi) \,  r_{1, j}] \, C_j(\mu)=1$ 
for the relevant operator.

We get that $G_{27}(\mu)[Q_1] = G_{27}(\mu)[Q_2]
= G_{27}(\mu)$ in Table \ref{tableBK} and they are therefore not listed
again. In addition all the other operators are
octet  so do not contribute to $G_{27}$. We also have $G_8^\prime(\mu)[Q_1]=0$,
the operator $Q_1$ only contributes via $B_K$-like contributions which
cannot have a contribution at $q^2=0$ for equal quark masses since
this type of contribution also produces $G_{27}$ where such terms
are forbidden. 
The approach to the chiral limit for the left-left 
current operators $Q_1$, $Q_2$, and
$Q_3$ is such that the $B_K$-like and Penguin-like contributions
are separately chiral invariant. For the left-right current
operator $Q_5$ this is not the case
and it is only the sum of the $B_K$-like and Penguin-like contributions
that vanishes for $q^2\to 0$ in the chiral limit. 
Notice that the results for small $\mu$ agree quite well
with the results just using CHPT, eq. (\ref{G8CHPT}),
but differ strongly for larger $\mu$.
The values (\ref{G8CHPT}) at $\mu=0$ correspond to the factorizable 
contribution.

We have also calculated the chiral logarithms that should be present in
these contributions. Subtracting them made the extraction of
the coefficient of $m_\pi^2$ to obtain $G_8^\prime$ numerically
much more convergent.
\begin{table}
\begin{center}
\begin{tabular}{|c|ccccccc|}
\hline
$\mu$ (GeV) & $G_8[Q_1]$&$G_8[Q_2]$&$G_8[Q_3]$& $G_8[Q_5]$ &
$G_8^\prime[Q_2]$ &  $G_8^\prime[Q_3]$&$G_8^\prime[Q_5]$\\[0.2cm]
\hline
0.0 & -0.667 & 1.000 & 0.000 & 0.000  & 0.000 & 0.000 &  0.000 \\
0.3 & -0.834 & 1.271 & 0.040 & -0.041 & 0.070 & 0.140 & -0.149 \\
0.4 & -0.930 & 1.425 & 0.060 & -0.109 & 0.128 & 0.256 & -0.297 \\
0.5 & -1.029 & 1.600 & 0.113 & -0.244 & 0.206 & 0.412 & -0.530 \\
0.6 & -1.130 & 1.779 & 0.168 & -0.460 & 0.298 & 0.596 & -0.868 \\
0.7 & -1.235 & 1.962 & 0.219 & -0.769 & 0.399 & 0.798 & -1.321 \\
0.8 & -1.347 & 2.145 & 0.249 & -1.178 & 0.501 & 1.002 & -1.908 \\
0.9 & -1.467 & 2.325 & 0.249 & -1.690 & 0.598 & 1.196 & -2.634 \\
1.0 & -1.597 & 2.498 & 0.205 & -2.308 & 0.681 & 1.362 & -3.504 \\
\hline
\end{tabular}
\end{center}
\caption{\label{tablecurrent} The results for the long-distance
contributions to $G_8(\mu)$ and $G_8^\prime(\mu)$ from $Q_1$ to $Q_5$ 
$[Q_4=Q_2-Q_1+Q_3]$ 
as calculated using the ENJL model via the two-point functions.}
\end{table}
The results for $G_8(\mu)[Q_3]$ can be obtained from isospin
relations from  $G_8(\mu)[Q_1]$  and $G_8(\mu)[Q_2]$.
The results for $G_8(\mu)[Q_5]$ come from  a large cancellation between
the values of $G_8-2G_8^\prime$ and $G_8^\prime$ and have a 
somewhat larger uncertainty than the others. 

It should be noticed that in all cases the $1/N_c$ corrections to the
matrix elements are substantial.

\subsection{The $Q_6$ Operator: Factorization Problem and Results}
\label{Q6discussion}

\begin{figure}
\begin{center}
\epsfig{file=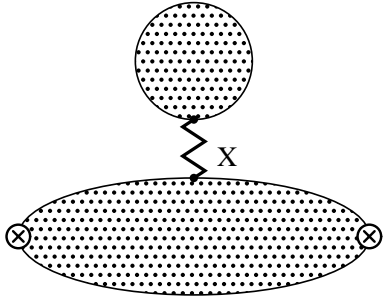,width=5cm}
\end{center}
\caption{\label{figfactQ6} The factorizable contribution for the $Q_6$
operator that is not well defined in the chiral limit. This contribution
is not present for current$\times$current operators.}
\end{figure}

After Fierzing, the $Q_6$ operator defined in (\ref{operators})
\ba
Q_6&\equiv&-2{\dis \sum_{q=u,d,s}} \, 
\left[\overline{s} \, (1+\gamma_5)\,q\right](x) \,
\left[\overline q \, (1-\gamma_5) \, d\right](x) 
\ea
gives both factorizable and  non-factorizable contributions to the off-shell
two-point functions to $K^0\to \pi^0$, $K^0\to\eta_8$, and  
$K^+\to \pi^+$. Here, we  study for definiteness the $K^+\to\pi^+$
two-point function, $\Pi^{K^+\pi^+}(q^2)$ of Eq. (\ref{two-point}).
The factorizable contributions  from $Q_6$ to this two-point function are
\ba
\label{Q6fac}
\Pi^{K^+\pi^+}_{Q_6 Fact}(q)&=& 2 C_{\Delta S=1} \, C_6(\mu) \left[ 
\langle 0| \overline  d d + \overline s s |0 \rangle  \, \, 
\Pi^{P_{K^-}S_{32}P_{\pi^+}}(0,q)  \right. \nonumber \\
&-& \left.  \Pi^P_{K^+K^+}(q) \,  \Pi^P_{\pi^+\pi^+}(q)\, \right]. 
\ea
Here $C_6(\mu)$ is the Wilson
coefficient of $Q_6$, $\Pi_P^{ii}(q)$ are two-point functions
\be
\Pi_P^{ii}(q) \equiv i \int {\rm d}^4 x \, e^{i q.x} \, 
\langle 0| T(P_{i}^\dagger(0) P_i(x)) | 0 \rangle
= -\left[ \frac{Z_i}{q^2-m_i^2}+ Z'_{i}\right],
\ee
with $P_{i}(x)$ the pseudo-scalar sources defined in (\ref{pseudosources}), 
and $\Pi^{P_{K^-}S_{32}P_{\pi^+}}(p,q)$ the three-point function
\ba
\Pi^{P_{K^-}S_{32}P_{\pi^+}}(p,q)&\equiv&
i^2 \int {\rm d}^4 x \, \int {\rm d}^4 y 
\, e^{i(q.x-p.y)} \, 
\langle 0| T(P_{K^-}(x)S_{32}(y) P_{\pi^+}(0)) | 0 \rangle \nonumber \\ 
\ea
with $S_{32}(y)$ the scalar source
\ba
S_{32}(y)\equiv - \left[ \overline s \, d \right] (y)\, .
\ea
The last term in Eq. (\ref{Q6fac}) corresponds to the diagram
shown in Fig. \ref{figfull}(a). The first term is a contribution
which is absent in the case of current$\times$current operators.
It is depicted in Fig. \ref{figfactQ6}. 

In octet symmetry,  to next-to-leading order we have \cite{GL}
\be
\frac{\displaystyle\langle 0| \overline  d d + \overline s s |0 \rangle}
{ -2 B_0 F_0^2} =
 1 + \frac{16}{F_0^2} \, (2 m_K^2+m_{\pi}^2)
L_{6} + \frac{4}{F_0^2} \, m_K^2 \, (2 L_8 + H_2 )
-  \frac{3}{2} \mu_\pi -3 \mu_K -\frac{5}{6} \mu_{\eta_8}
\ee
for the one-point function and
\ba
\lefteqn{\Pi^{P_{K^-}S_{32}P_{\pi^+}}(0,q)=
- \frac{\sqrt{Z_{K^+}Z_{\pi^+}}}{(q^2-m_K^2)(q^2-m_{\pi}^2)} 
\, B_0}&& \nonumber \\ &\times&
 \left[ 1+ \frac{8}{F_0^2} \, (2 m_K^2+m_{\pi}^2) 
(2 L_{6} - L_4) \right.
-  \frac{4}{F_0^2} \, (2q^2+m_K^2+m_\pi^2) \,  L_5  
+ \frac{32}{F_0^2} \, q^2  \, L_8 \nonumber \\ &&
\!\!\!\!\!\!\!- {1\over6} \frac{q^2}{16 \pi^2 F_0^2} \left [
\ln\left(\frac{m_K^2}{\nu^2}\right)- {3\over2(m_K^2-m_\pi^2)} 
\left( m_\pi^2 \ln\left({m_\pi^2\over m_K^2}\right) +
m_{\eta_8}^2 \ln\left({m_{\eta_8}^2\over m_K^2}\right) 
\right) \right] \nonumber \\ &&
+ {1\over2} \mu_\pi + {1\over6} \mu_{\eta_8} +
{5 \over 6} \mu_K + \frac{5}{12} \, 
\frac{m_\pi^2}{16 \pi^2 F_0^2} \ln\left(\frac{m_K^2}{\nu^2}\right)
\nonumber \\ &&
- \left.  \frac{m_K^2+m_\pi^2}{16 \pi^2 F_0^2} \, 
{1\over8(m_K^2-m_\pi^2)} \left[
m_{\eta_8}^2 \ln\left({m_{\eta_8}^2\over m_K^2}\right) 
- 3 m_{\pi}^2 \ln\left({m_{\pi}^2\over m_K^2}\right) 
\right] \right]
\ea
for the three-point function.
Here and in the remainder the constants $L_i$ are defined at a scale $\nu$,
$L_i\equiv L_i^r(\nu)$.
and $\mu_i=\ln(m_i/\nu)/(16\pi^2)$ for $i=\pi,K,\eta_8$.

At next-to-leading order, the
 expressions for the two-point functions were given
for the octet symmetry case in \cite{BPP98}. So the second part in 
(\ref{Q6fac}) can be written as
\ba
\lefteqn{\Pi^P_{K^+K^+}(q) \, \Pi^P_{\pi^+\pi^+}(q) =
\frac{\sqrt{Z_{K^+}Z_{\pi^+}}}{(q^2-m_K^2)(q^2-m_{\pi}^2)} 
\, 2 F_0^2 B_0^2 }&& \nonumber \\ &\times&
 \left[ 1+ \frac{8}{F_0^2} 
(2 m_K^2 +m_\pi^2)  (4 L_6 -L_4) \right.
+ \frac{4}{F_0^2} (m_\pi^2+m_K^2) (4 L_8 -L_5)
\nonumber\\&& +
\frac{4}{F_0^2} (2 q^2 - m_\pi^2-m_K^2) (2 L_8 -H_2)
- \left. 
\frac{7}{4} \mu_\pi - \frac{5}{2} \mu_K -\frac{5}{12} \mu_{\eta_8}
 \right]
\ea

Therefore this order,  in octet symmetry, the factorizable contributions
to $\Pi^{K^+\pi^+}(q)$ from $Q_6$ are
\ba
\lefteqn{\Pi^{K^+\pi^+}_{Q_6, Fact}(q) =
-\frac{\sqrt{Z_{K^+}Z_{\pi^+}}}{(q^2-m_K^2)(q^2-m_{\pi}^2)} 
\, C_{\Delta S=1} \, C_6(\mu) 
\, 16  B_0^2(\mu) }\nonumber&& \\&\times& 
\Bigg\{ 2 q^2 \left\{L_5 -(2 L_8 + H_2)\right\} +
 m _\pi^2 (2 L_8 + H_2)  
-{1\over12 } \mu_K +{1\over16} \mu_{\eta_8} - {3\over16} \mu_\pi
\nonumber \\ &&\hskip-0.8cm
+ {1\over24} \frac{q^2}{16 \pi^2 F_0^2} \left [
\ln\left(\frac{m_K^2}{\nu^2}\right)- {3\over2(m_K^2-m_\pi^2)} 
\left( m_\pi^2 \ln\left({m_\pi^2\over m_K^2}\right) +
m_{\eta_8}^2 \ln\left({m_{\eta_8}^2\over m_K^2}\right) 
\right) \right] \nonumber \\ &&
-  \frac{5}{48} \, 
\frac{m_\pi^2}{16 \pi^2 F_0^2} \ln\left(\frac{m_K^2}{\nu^2}\right)
\nonumber \\ &&
+   \frac{m_K^2+m_\pi^2}{16 \pi^2 F_0^2} \, 
{1\over32(m_K^2-m_\pi^2)} \left[
m_{\eta_8}^2 \ln\left({m_{\eta_8}^2\over m_K^2}\right) 
- 3 m_{\pi}^2 \ln\left({m_{\pi}^2\over m_K^2}\right) 
\right] \Bigg\}
\ea
As it is well known the order $p^0$ contribution from $Q_6$ vanishes
\cite{CFG86}
and the first non-trivial contribution from this operator
is of order\footnote{The  order $p^2$ chiral logs were called
order $p^0/N_c$ contributions in \cite{GKPSB}.} \, $p^2$. 
This happens here as an exact
cancellation between the two types of factorizable
contributions at order $p^0$. As a result there is a very large 
cancellation between the two types of factorizable contributions
at order $p^2$.
We get 
\be
\label{G8Q6fact}
G_8\Bigg|_{Q_6,\mbox{Fact}} = - \left[ {5\over 3} \right] \, 16 \,  
C_6(\mu) \frac{B_0^2(\mu)}{F_0^2} \, \left[ L_5 
- {3\over 16} \frac{1}{16 \pi^2 }
\left[ 2 \ln\left(\frac{m_L}{\nu}\right) + 1 \right]
\right] \nonumber \\
\ee
and 
\be
\label{G8pQ6fact}
G_8^{\prime}\Bigg|_{Q_6,\mbox{Fact}} =
- \left[ {5\over 3} \right] \, 8 \,  
C_6(\mu) \frac{B_0^2(\mu)}{F_0^2} \, \left[ (2L_8+H_2)
-\frac{5}{24} \, \frac{1}{16 \pi^2 } 
\left[ 2 \ln\left(\frac{m_L}{\nu}\right) + 1\right] \right]
\ee
The mass $m_L$ above has to be understood as an infrared cut-off
as we have done the chiral limit $m_L = m_\pi=m_K=m_{\eta_8} \to 0$.
The factorizable contribution  to $G_8$ and G$_8'$ from $Q_6$ 
is therefore not well defined. It has an infrared divergence. 
The divergence is related to the divergence in the
pion scalar radius in the chiral limit. Since $Q_6$ is an 
$8_L\times 1_R$
operator we know from CHPT in the non-leptonic sector that to lowest order
in the counting there, no infrared divergences are present in the
two-point function $\Pi^{K^+\pi^+}(q^2)$. These infrared
divergences are therefore
spurious and must be cancelled by another contribution.
The only possibility is that it
cancels out with the non-factorizable contribution also coming from $Q_6$.
We will see below that this is indeed the case.
Notice also  that since $G_8$ and $G_8'$  are $O(p^2)$ couplings, 
Eqs. (\ref{G8Q6fact}) and (\ref{G8pQ6fact})
are  exact for the factorizable contributions.

Unfortunately, the non-factorizable contributions can only be calculated
at present in a model dependent way. In the $1/N_c$ expansion,
the infrared divergent part of $G_8$ and $G_8^\prime$
can in fact be calculated analytically using the $O(p^2)$ CHPT Lagrangian.
We can therefore subtract it.  It follows from the diagrams shown
in Fig. \ref{figBK}, (b),(c), (e),
and (f) and by using CHPT for the $X$-boson vertices which is valid for
small $\mu$. For equal masses $m_K^2 = m_\pi^2 = m_{\eta_8}^2 = m_L^2$ 
we obtain
\ba
\label{Q6NF-CHPT}
\lefteqn{\Pi^{K^+\pi^+}(q^2) = \frac{2B_0^2 F_0^2}{(q^2-m_K^2)(q^2-m_\pi^2)}
C_{\Delta S=1} C_6(\mu)4B_0^2(\mu)}&&\nonumber\\&\times&
\Bigg\{-\frac{1}{6}(q^2-5m_L^2)\int^\mu 
\frac{d^4r_E}{(2\pi)^4}\frac{1}{(r_E^2+m_L^2)^2} \nonumber\\&&
-\frac{5}{6} \int^\mu 
\frac{d^4r_E}{(2\pi)^4}\left[\frac{1}{((r_E+q_E)^2+m_L^2)}
-\frac{1}{(r_E^2+m_L^2)}\right] \Bigg\}\, . \nonumber \\
\ea
The non-factorizable (NF) part above in the limit $m_L\to0$ leads to
\be
\label{G8Q6NF-IR}
G_8{\Bigg|_{\mbox{$Q_6$,NF-$O(p^2)$}}} 
= - \left[ {5\over 3} \right] \, 16 \,  
C_6(\mu) \frac{B_0^2(\mu)}{F_0^2} \, \frac{3}{16}\, \frac{1}{16\pi^2}
\left[2 \ln\left(\frac{m_L}{\mu}\right)+\frac{13}{18}\right]
\ee
and 
\be
\label{G8pQ6NF-IR}
G_8^{\prime}{\Bigg|_{\mbox{$Q_6$,NF-$O(p^2)$}}} = 
- \left[ {5\over 3} \right] \, 8 \,  
C_6(\mu) \frac{B_0^2(\mu)}{F_0^2} \, \frac{5}{24} \, \frac{1}{16\pi^2}
\left[2 \ln\left(\frac{m_L}{\mu}\right)+1\right]
\ee

There is a very large cancellation between the factorizable
parts in (\ref{G8Q6fact}) and (\ref{G8pQ6fact}) and the non-factorizable 
part in (\ref{G8Q6NF-IR}) and (\ref{G8pQ6NF-IR}) 
both for the IR divergent part and for the large $1/N_c$ constant
part.  Summing up the exact factorizable result and the infrared divergent
non-factorizable part we get
\be
\label{G8Q6}
G_8\Bigg|_{\mbox{$Q_6$, $O(p^2)$}} = - \left[ {5\over 3} \right] \, 16 \,  
C_6(\mu) \frac{B_0^2(\mu)}{F_0^2} \, \left[ L_5(\nu) 
-  \frac{1}{16 \pi^2 }\left(
{3\over 8} \ln\left(\frac{\mu}{\nu}\right)
+\frac{5}{96}\right)\right]
\ee
and 
\be
\label{G8pQ6}
G_8^{\prime}\Bigg|_{\mbox{$Q_6$, $O(p^2)$}} =
- \left[ {5\over 3} \right] \, 8 \,  
C_6(\mu) \frac{B_0^2(\mu)}{F_0^2} \, \left[ (2L_8+H_2)(\nu)
-\frac{5}{12} \, \frac{1}{16 \pi^2 } 
 \ln\left(\frac{\mu}{\nu}\right)  \right]
\ee
It is then  a 
non-trivial check of the validity of the model used that the 
 non-factorizable part indeed contains the correct
infrared logarithms needed to cancel the factorizable ones.
 The ENJL model used here does.

Notice in (\ref{G8Q6}) and  (\ref{G8pQ6}) all the dependence
 on the IR scale , $m_L^2$, drops out as it should
and the scale in the logarithm becomes $\ln(\mu/\nu)$.
So in the chiral limit and next-to-leading
in $1/N_c$ , the scale dependence on the short-distance 
scale gets  compared to the scale where the CHPT constants are defined.

The result above shows that at least the $B_6$ parameter defined
as usual as the ratio of the non-factorizable contributions over the 
vacuum saturation result (VSA) is not well defined.
It is therefore necessary  to give another definition for this $B$
parameter.  The cancellation of the infrared divergence found
here is probably also the source for the large cancellations found
between the factorizable and non-factorizable contributions
in earlier work. Notice also that the $1/N_c$ finite term 
in (\ref{G8Q6fact}) is {\em larger} than the leading in 
$1/N_c$ result and
with {\em opposite} sign. It is clear  that it can be
dangerous not to have and
analytical cancellation of both the IR divergent  part 
and the $1/N_c$ constant as we have. This can explain also
some discrepancies for the $B_6$ parameter results in the literature, 
 $B_6$ is  just not well defined.

The way we treat our results is that we remove the exact infrared
logarithm from our ENJL calculation by adding equations (\ref{G8Q6fact})
and 
(\ref{G8pQ6fact}) which are exact and model independent to the ENJL results.
In this way we also remove the IR divergence of the non-factorizable
part exactly. We chose the reference scale $\mu = M_\rho$
to do the subtraction. We generate the mass $m_L^2$ 
by putting small current quark masses.
The remaining factorizable factor, i.e.
the part from the constants $L_5$, $L_8$, and $H_2$ are then evaluated
at a scale $\nu = M_\rho$. 
This corresponds for the leading in  $1/N_c$ contribution 
to $G_8$ and $G_8'$ from $Q_6$ 
\be
\label{numfact}
G_8^{ENJL}{\Bigg|_{\mbox{$N_c$}}} = (-38\pm8) \, C_6(\mu)
\quad{\rm and} \quad 
G_8^{\prime ENJL}{\Bigg|_{\mbox{$N_c$}}} = (-9\pm14) \, C_6(\mu)
\ee
using
\ba
 L_5(M_\rho)&=&(1.4\pm0.3)\cdot 10^{-3} \nonumber \\ 
 (2L_8+H_2)(M_\rho) &=& (0.7\pm1.1)\cdot10^{-3} \, .
\ea
We have used here the value of $B_0$ and $F_0$ from
the ENJL model. The value of $2L_8+H_2$ is derived from the
canonical value for $L_8(M_\rho) = (0.9\pm0.3)\cdot10^{-3}$ and the
value for $(2L_8-H_2)(M_\rho)=
(2.9\pm1.0)\cdot10^{-3}$ from \cite{BPR}.
The large error for  $G_8^{' ENJL}$ in (\ref{numfact}) is because of the
large cancellation in the value for $2L_8+H_2$.
Notice that the size of the subtracted terms in 
$G_8^{ENJL}$ is about 
$+40 \, C_6(\mu)$ for $m_L^2=m_\pi m_K$ and varies very fast
with $m_L$.  

Our calculation agrees with the one of \cite{GKPSB} when the
appropriate identifications are made. The large cancellation between the
factorizable and non-factorizable parts where also observed there.
They were however not identified as an exact cancellation of infrared
divergences. In fact, at the order the calculation was done
in \cite{GKPSB} the cancellation of the $1/N_c$ factorizable
and non-factorizable pieces is very large, and 
 in their language\footnote{As we said
$B_6$ is not well defined. We come back to this question in Section 
\ref{conclusions}.}  one should get $B_6$ very near to one. 
They get indeed $B_6$ very close to one.

The non-factorizable non-divergent part has 
corrections from higher order terms in the chiral
Lagrangian which we  calculate numerically using the ENJL model. 
We have included  them  and  these give therefore the numerical differences
between our results and the ones in \cite{GKPSB}.

Before we present the results for $G_8(\mu)$ 
and $G_8'(\mu)$ from $Q_6$ from our ENJL calculation we need to 
include one additional remark.
The vector and axial-vector currents used in the previous section are
uniquely identified both in the ENJL model and in QCD. There is
however no guarantee as remarked in \cite{BP94} that the same is true
for the scalar and pseudo-scalar densities. Here we renormalize
the ENJL scalar $S(x)$  and pseudo-scalar $P(x)$ densities
 by the values of the quark condensates in the chiral limit:
\be
\label{renormB0}
S_{\mbox{ENJL}} = S_{\mbox{QCD}}(\mu)
\frac{\langle\bar{q}q\rangle_{\mbox{ENJL}}} 
{\langle\bar{q}q\rangle_{\mbox{QCD}}(\mu)}.
\ee
There is an analogous equation for the pseudo-scalar density.
This factor should be remembered when using the Wilson 
coefficients from our results. 
The values we have used are $B_0^{QCD}(1 {\rm GeV})=(1.75\pm0.40)$ GeV
in the $\overline{MS}$ scheme \cite{BPR,DN98}, and $B_0^{ENJL}=2.80$ GeV
\cite{BBR}. We have also included the QCD scale dependence of
the $B_0$ parameter to two-loops.
We show in Table \ref{tableQ6} the results
for $G_8(\mu)[Q_6]$ and $G_8^\prime(\mu)[Q_6]$ without the
renormalization factor of Eq. (\ref{renormB0}), columns labelled ENJL, and
including the
renormalization factor of Eq. (\ref{renormB0}) both to one-loop,
columns labelled $^{(1)}$,
and two-loops in QCD , columns labelled $^{(2)}$.
Notice $B_0(\mu)=-\langle\bar{q}q\rangle(\mu)/F_0^2$ and this factor
is responsible for most of the running of $Q_6$\cite{EdR89}.

\begin{table}
\begin{center}
\begin{tabular}{|c|cccccc|}
\hline
$\mu$ (GeV) &$G_8[Q_6]$&
$G_8^\prime[Q_6]$&
$G_8[Q_6]$&
$G_8^\prime[Q_6]$& 
$G_8[Q_6]$ &
$G_8^\prime[Q_6]$\\[0.2cm]
 & ENJL & ENJL & $^{(1)}$ &$^{(1)}$&$^{(2)}$&$^{(2)}$\\
\hline
0.3 & -118 & -69 &&&&  \\
0.4 & -103 & -53 &&&&  \\
0.5 & -93  & -41 & -21.1  & -9.3 & -6.4 & -2.8  \\
0.6 & -88  & -32 & -23.9  & -8.7 & -14.7 & -5.3 \\
0.7 & -84  & -25 & -25.9  & -7.7 & -20.1 &-6.0   \\
0.8 & -82  & -20 & -27.9  & -6.8 & -24.5 & -6.0 \\
0.9 & -82  & -17 & -30.0  & -6.2 & -28.4 & -5.9 \\
1.0 & -83  & -15 & -32.4  & -5.9 & -32.4 &-5.9   \\
\hline
\end{tabular}
\end{center}
\caption{\label{tableQ6} Results for the long-distance contributions
to $G_8$ and $G_8^\prime$ from $Q_6$ as calculated using the ENJL model via
the two-point functions for the non-factorizable part and adding
the model independent factorizable part in (\ref{G8Q6fact}) and
(\ref{G8pQ6fact}).
The last 4 columns  include the renormalization of scalar and
pseudo-scalar densities to one-loop $^{(1)}$ and two-loops $^{(2)}$
in QCD. The short-distance
anomalous dimensions for $B_0(\mu)$ at scales below 0.5 GeV
blows up.}   
\end{table}

\section{The Order $p^2$ Full $\Delta S=1$ Couplings}
\label{fullresults}

We use here the results of \cite{two-loops} and \cite{one-loop}
for the $\Delta S=1$ QCD anomalous dimensions to one- and two-loops
respectively to obtain final values. 
The solution for the  Wilson coefficients are given in 
\cite{two-loops,BurasReviews} at two-loops
using an expansion in $\alpha_s$. Whenever
the values of  $\Lambda_{QCD}$ are needed in the
$\overline{MS}$ scheme with three flavours 
we use the expanded in $\alpha_s$ formulae \cite{PDG} from
$\alpha_s(M_\tau)=0.334\pm0.006$ 
with $M_\tau=1.77705 \pm 0.00030$ GeV \cite{PDG}
and get $\Lambda^{(1)}_{QCD}=$ 220 MeV to one-loop
and $\Lambda^{(2)}_{QCD}=$ 400 MeV to two-loops.  The
values of the Wilson coefficients we use 
for $\Delta S=1$ \cite{two-loops,BurasReviews} 
and for $\Delta S=2$ \cite{S=2twoloops}
are in the Appendix.  
We also include there the scheme dependent constants $r_1$ needed for the
two-loops short-distance running in the NDR scheme we use.

We now show in Tables \ref{resultsg27g8g8p} and \ref{resultsg27g8g8p2}
the results
for the coefficients $G_{27}$, $G_8$ and $G_8^\prime$.
The numbers in brackets refer to keeping only $Q_1$, $Q_2$, and $Q_6$.
Most of the difference is due to $Q_4$.

The matching for the one-loop running of the Wilson coefficients is very good.
We obtain a value of $G_8\approx4.3$ and $G_8^\prime\approx0.8$.
 If we look inside the numbers,
for $G_8$ the contribution via $Q_1$ is fairly constant over the whole
range but there is a distinct shift from $Q_2$ to $Q_6$ for lower
values of $\mu$. The operator $Q_2$ remains the most important over the
entire range of $\mu$ considered. For $G_8^\prime$ similar comments
apply except that $Q_1$ doesn't contribute. 
Typically $G_{27}$ is somewhat low compared to the experimental number
and we have not as good matching as in the octet sector.
Notice though  that it gets  somewhat more stable 
in the range between 0.5 and 0.8 GeV as one expects from
the validity of the low-energy model.

When two-loop running is taken into account in the
NDR scheme the numbers do not change so much. The effect of the
$r_1$ constants in this scheme 
is however very large and causes a significant
shift in the numbers.

The numbers for the octet case are somewhat stable
in the range $\mu=0.8$ to $1.0$ GeV but there is where the ENJL model
is expected to start deviating from the true behaviour.
\begin{table}
\begin{center}
\begin{tabular}{|c|ccc|}
\hline
$\mu$ (GeV)& $G_{27}$ & $G_{8}$ & $G_8^\prime$\\
\hline
0.5 & 0.399 & 4.45 (4.55) & 0.739 (0.761)\\
0.6 & 0.351 & 4.26 (4.34) & 0.686 (0.710)\\
0.7 & 0.291 & 4.21 (4.28) & 0.703 (0.727)\\
0.8 & 0.221 & 4.25 (4.30) & 0.767 (0.789)\\
0.9 & 0.141 & 4.33 (4.37) & 0.847 (0.866)\\
1.0 & 0.050 & 4.44 (4.46) & 0.923 (0.935)\\
\hline
\end{tabular}
\end{center}
\caption{\label{resultsg27g8g8p} The final results for the three
$O(p^2)$ couplings using the one-loop Wilson coefficients. The numbers
in brackets refer to using $Q_1$, $Q_2$, and $Q_6$ only.}
\end{table}
\begin{table}
\begin{center}
\begin{tabular}{|c|ccc|}
\hline
$\mu$ (GeV) & $G_{27}$ & $G_{8}$ & $G_8^\prime$\\
\hline
0.5 & 0.182 & 11.20 (12.4)& 1.60 (1.75)\\
0.6 & 0.249 & 7.30 (7.8) & 1.13 (1.22)\\
0.7 & 0.230 & 6.30 (6.6) & 0.99 (1.10)\\
0.8 & 0.184 & 5.88 (6.2) & 0.97 (1.08)\\
0.9 & 0.121 & 5.73 (5.9) & 0.99 (1.11)\\
1.0 & 0.044 & 5.61 (5.8) & 1.03 (1.14)\\
\hline
\end{tabular}
\end{center}
\caption{\label{resultsg27g8g8p2} The final results for the three
$O(p^2)$ couplings using the two-loop Wilson coefficients
with the inclusion of the $r_1$ factors. The numbers
in brackets refer to using $Q_1$, $Q_2$, and $Q_6$ only.}
\end{table}

Notice that at large $N_c$, $G_8$ and $G_{27}$
are both 1. Adding $1/N_c$ corrections
 $G_{27}$ decreases by a non-negligible factor 
around two to three, while the $G_8$ coupling gets 
enhanced up to $G_8=6.2 \pm0.7$.
The short-distance enhancement is almost a factor
of two for the whole range of $\mu$. The rest of the
enhancement, namely a factor two to three is mainly due to the
large value of the long-distance contribution
to the Penguin-like coupling $c$. 
The bulk of the long distance part enhancement of the coupling $c$
comes from  $Q_2$ and $Q_6$. There is also a
small contribution to $G_8$ in the right direction from the $B_K$-like
coupling $b$ from both $Q_2$ and $Q_1$.

The final results for the ratio $|A_0/A_2|$ at 
$O(p^2)$ (\ref{ratioA0A2p2}) are in Table \ref{resultsA0/A2}.
The stability we get for the one-loop short-distance is not bad,
and  there is some minimum  around 0.7 GeV for the two-loop running. 
We get in general too large values for this ratio compared 
to the experimental 16.4 value (\ref{I=1/2p2}) due to the 
somewhat small value of $G_{27}$ we get. 
\begin{table}
\begin{center}
\begin{tabular}{|c|cc|}
\hline
$\mu$ (GeV) & One-Loop & Two-Loops \\
\hline
0.5 & 14.3 & 78.5 \\
0.6 & 15.6 & 37.5  \\
0.7 & 18.6 & 35.0  \\
0.8 & 24.6 & 40.8  \\
0.9 & 39.2 & 60.1  \\
1.0 & 113.2 & 162.4  \\
\hline
\end{tabular}
\end{center}
\caption{\label{resultsA0/A2} The final results for the 
ratio $|A_0/A_2|$ to $O(p^2)$ using the one-loop short-distance
running and the full scheme independent two-loops short-distance running.}
\end{table}

\begin{figure}
\begin{center}
\epsfig{file=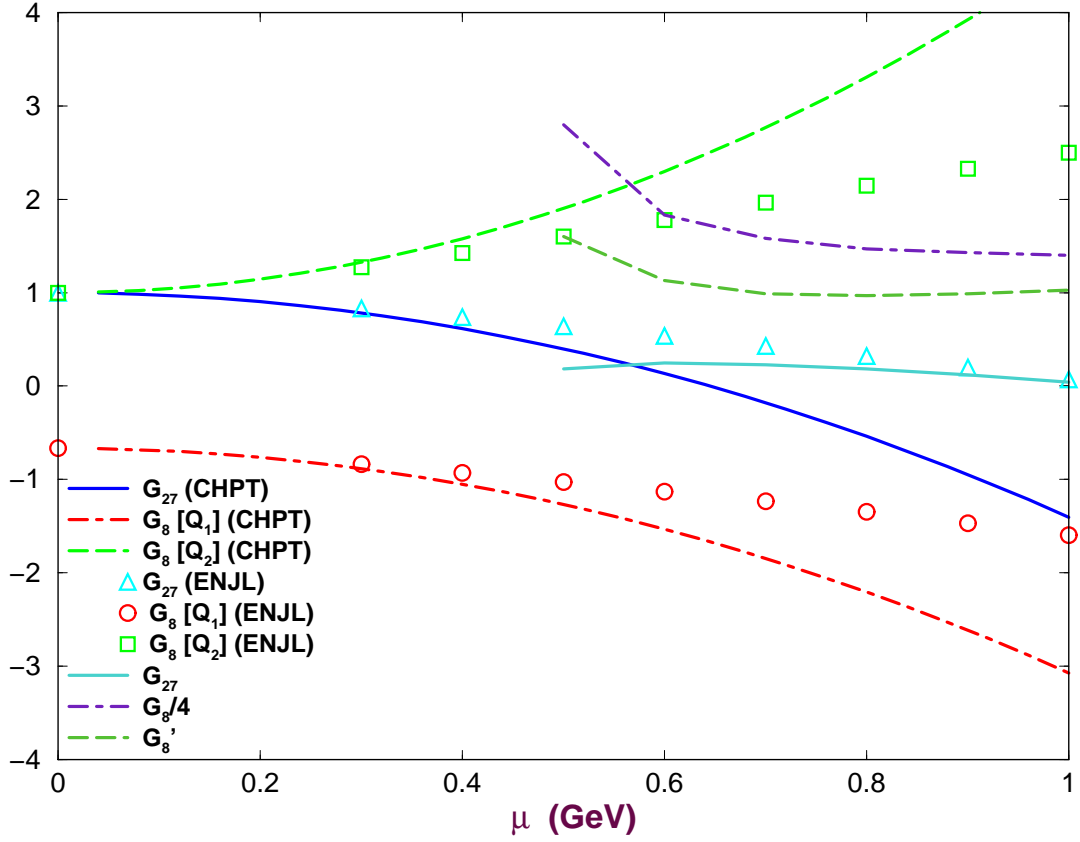,width=13cm,angle=-90}
\end{center}
\caption{\label{figmatch} The improvement of the behaviour with $\mu$
of several quantities. Shown are the lowest order result, ENJL result,
and when short-distance running is added.}
\end{figure}

In order to show the improvement with previous results and the quality
of the matching we have shown in Figure \ref{figmatch} for $G_{27}(\mu)$
the lowest order result Eq. (\ref{G8CHPT}), the ENJL result for the
same quantity and the final result for $G_{27}$ with the two-loop
short distance included. We have similarly plotted $G_8[Q_1](\mu)$ and
$G_8[Q_2](\mu)$ both from the lowest order result Eq. (\ref{G8CHPT})
and from the ENJL model. We also showed the full result for $G_8^\prime$
when the two-loop running is included properly. Similar improvements of
Eq. (\ref{G8CHPT}) and (\ref{G8Q6}) can be seen by plotting the other
results with the corresponding ones from Tables \ref{tablecurrent},
\ref{tableQ6}, and \ref{resultsg27g8g8p2}.

In summary, the results we get for $G_8$, $G_{27}$, and $G_8'$ are 
\ba
\label{finalnumbers}
4.3 < &G_8 &< 7.5 \nonumber\\
0.8 < & G_8^\prime & < 1.1\nonumber\\
0.25 < &G_{27}& < 0.40
\ea 
The bounds have been chosen by looking at both the one-loop and two-loop
results in the stability regions in 
Tables \ref{resultsg27g8g8p} and \ref{resultsg27g8g8p2}.
{}From (\ref{finalnumbers}) we can extract the values
\ba
-0.75 &< b<& -0.50 \nonumber \\
1.7 &< c <& 3.7 
\ea
and we have fixed $a=1$ as explained before. 
For the $\Delta I=1/2$ rule we get
\ba
15 < \left| \frac{A_0}{A_2} \right|^{(2)} < 40
\ea
to order $p^2$.

We get a huge enhancement due to the $c$-coupling, it is therefore
interesting what do other calculations predict for this coupling.
One model where this coupling can be easily extracted is 
 the effective action approach \cite{PdeR91}. 
To order $1/N_c$ one gets  \cite{PdeR91,BP93}
\ba
c&=& C_2(\mu) - 1 + \Re e \, C_4(\mu) + 
C_2(\mu) \, \frac{4 \pi\alpha_s(\mu)}{N_c}
\, (2H_1+L_{10})(\mu) \nonumber \\
&-& 16 \frac{B_0^2(\mu)}{F_0^2}\,
  L_5(\mu) \left[ \Re e \, C_6(\mu) + C_2(\mu) \,
 \frac{4 \pi\alpha_s(\mu)}{N_c}
\, (2H_1+L_{10})(\mu) \right] \nonumber \\ &+& O(1/N_c^2)
\ea
with $\mu=M_\rho$, $\alpha_s(M_\rho)=0.70$, $B_0(M_\rho)=1.4$ GeV, 
and $(2H_1+L_{10})(M_\rho)= -0.015$ \cite{BBR}, 
we get 
\ba
c&=&0.95 \pm 0.40  \, .
\ea
The reason why $c$ is  smaller than the present work  results
is that the long-distance mixing between $Q_2$ and $Q_6$
is not well treated in this model. In fact this contribution
is model dependent already at $O(1/N_c)$. For instance, it  appears
in terms of the short-distance value $\alpha_s(M_\rho)$.
It is clear that at such scales  one has to treat the long 
distance contributions in a hadronic model and the $\alpha_s(M_\rho)$
above will appear enhanced. Nevertheless, the extra contribution
to $c$ coming from the operator $Q_2$ \cite{PdeR91,BP93}
 both from short-distance origin, namely the term $C_2(\mu)-1$,
and from long-distance origin, namely  the part proportional to
$2H_1+L_{10}$, give some insight on the potentially large value of $c$.

We cannot easily compare our result with those of \cite{Trieste},
their method of calculating the low energy part has no obvious connection 
to the short-distance evolution and their results cannot be directly
compared to ours. The results from the lattice \cite{lattice}
are at rather high values of the quark masses and can thus also not be
simply compared to our results. 

As stated above, we agree with the calculations of \cite{GKPSB}
for low values of the scale $\mu$ where we should agree but deviate
significantly at higher scales. The earlier Dortmund group
results \cite{Dortmund} are thus also expected to have significant
corrections. The attempts at calculating via more inclusive modes
\cite{PdeR}
have very large QCD corrections\cite{PdeR91,Pich}. We see the remnant
of this in the large corrections from the $r_1$ terms, see 
Appendix \ref{AppA}. The short-distance factors are in fact
one of the bigger remaining sources of uncertainty.

\section{Results and Conclusions}
\label{conclusions}

The main results of this paper are the results for the $O(p^2)$ couplings
$G_8$, $G_{27}$, and $G_8^\prime$ as a function of cut-off 
$\mu$ for the various
operators $Q_j$,  $j=1,\cdots, 6$, as given 
in Tables \ref{tableBK}, \ref{tablecurrent}, and \ref{tableQ6}.
In addition we have corrected our earlier results for $B_K(\mu)$ for
the routing problem as described in Section \ref{routing} and presented
those in Table \ref{tableBK} as well.

The other main result of this paper is the observation that
in the chiral limit the factorizable contribution from $Q_6$ is not well
defined due to an infrared divergence
and we expect that similar problems will show up for the
current-current operators when we try to calculate higher order coefficients
in the weak chiral perturbation theory Lagrangian. We showed that
the total contribution of $Q_6$ obtained after adding the non-factorizable
and factorizable parts is however well defined. We also expect that
he same solution will hold for coefficients of higher order operators
in the chiral lagrangian. A corollary of this observation is
that the use of $B$-factors in the chiral limit
as is common in other treatments of weak
non-leptonic operators is not possible in the way they are defined, namely
the whole result normalized to the  VSA result.

One could use the leading result in $1/N_c$ as an appropriate starting
point for normalizing the $B_6$-parameter in the chiral limit,
this is keeping only the $L_5$, $L_8$, and $H_2$ terms but this is difficult
to implement for lattice gauge theory calculations.
In fact, what in practice people have used \cite{BurasReviews,GKPSB}
for the VSA, i.e. the 
factorizable part of  $Q_6$,  has been  just the large $N_c$ part.
Of course, this  is not in agreement with what is done with other $B$
parameters for current $\times$ current operators
like  $B_K$ where the $1/N_c$ factorizable  part
is always included in the VSA result.
 After the problems we encountered and the
importance of the $B$ parameters to  normalize results from different
techniques,
we believe a new consistent definition 
of the  $B$ parameters should be looked for or just abandon the use
of $B$ parameters  and quote  matrix elements values.
We also emphasize that
caution should be taken when combining results from different methods for
the factorizable and non-factorizable contributions.

When we combine our main results with the Wilson coefficients at one-loop
we get nice stable results. Using the Wilson coefficients at two-loops with
the inclusion of the $r_1$ factors which as we argued in Section \ref{Xboson}
is necessary we obtain relatively stable values for $G_8$ and the coefficient
of the weak mass term $G_8'$ with
\ba
4.3 < &G_8 &< 7.5 \nonumber\\
0.8 < & G_8^\prime & < 1.1\nonumber\\
0.25 < &G_{27}& < 0.40
\ea
The main uncertainty here is in fact coming from the short-distance
coefficients for the octet case and from the long-distance for the 
27-plet case. For the $G_{27}$ coupling we obtain a somewhat small
value compared to the experimental one. This translates into the
following results for the $\Delta I=1/2$ rule in the chiral limit
\be
15< \left| \frac{A_0}{A_2} \right|^{(2)} < 40\,.
\ee 
These results are somewhat large.
Nevertheless, we would like to emphasize that we have obtained these 
results from 
a next-to-leading in $1/N_c$ long-distance calculation and we have
passed from the large $N_c$ result $| A_0/A_2 |_{N_c}=\sqrt 2$ 
to values around 20 to 35. One
can certainly expect non-negligible $1/N_c^2$ corrections to our 
results but the huge enhancement is there. We would also like to stress
that we have no free input in our calculation. All parameters have been
determined from elsewhere.

{}From the results above  we have also obtained the couplings
$b=G_{27}-1$ and $c=(3G_8+2G_{27})/5-1$
\ba
-0.75 &< b<& -0.50 \nonumber \\
1.7 &< c <& 3.7 \, . 
\ea
Here is then one of our main results, the $\Delta I=1/2$
rule enhancement comes from  the 
Penguin-like topologies $(c)$ in Figure \ref{figfull}, both from
$Q_2$ which dominates for high values of $\mu$ 
and from $Q_6$ which dominates for small values of $\mu$.

In addition we obtain a value for chiral limit value 
$\hat B_K^\chi$ as defined in Eq. (\ref{defBKhat})
\be
0.25<\hat B_K^\chi <0.40 
\ee
and the value for the $\hat B_K$ parameter
in the real case
\be
0.59<\hat B_K < 0.79 \, . 
\ee
 These two results confirm the ones in \cite{BPBK}.
Notice that the different short-distance contribution 
{}from $M_W$ until the charm quark mass  to $G_{27}$ and
 $\hat B_K^\chi$  has produced
\be
\frac{\hat B_K^\chi}{G_{27}} \simeq 1.1 \,  
\ee
instead of $3/4$.

So we have obtained quite good matching for $G_8$, $G_8^\prime$,
and $\hat B_K$ for values of $\mu$ around $0.7-1.04$ GeV
and for $G_{27}$ for values of $\mu$ around $0.6-0.8$ GeV.
We obtained values
for the three parameters of order $p^2$ not too far from the
experimental ones and a quantitative
understanding of the origin of the $\Delta I=1/2$ enhancement.
Notice that the values of the cut-off we use to predict
our results are not extremely low as in other $1/N_c$ approaches,
still one would like the matching region to be larger and
for somewhat larger values of the cut-off.

\acknowledgments

 This work was partially supported by the European Union
TMR Network $EURODAPHNE$ (Contract No ERBFMX-CT98-0169) and by the 
Swedish Science Foundation. 
The work of J.P. was supported in part
by CICYT (Spain) and by Junta de Andaluc\'{\i}a under Grants Nos.
AEN-96/1672 and FQM-101 respectively. J.P. also likes to thank the
  CERN Theory Division and the Department of Theoretical Physics at Lund
University  (Sweden) where part of his work was done for hospitality.
We thank Elisabetta Pallante for participation in the early parts of this work
and Eduardo de Rafael for discussions

\appendix

\section{$\Delta S=1$ and $\Delta S=2$ Wilson Coefficients}

\label{AppA}

In this section we give the numerical values of the Wilson coefficients
for the basis of $\Delta S=1$ operators in
(\ref{q2}), (\ref{q1}), (\ref{operators}),
 and for the $\Delta S=2$ operator in (\ref{qS2}). We give them 
for the relevant values of the renormalization
scale. We have extensively used the formulae in \cite{BurasReviews}.

In all cases we have used
$\alpha_s(M_W) = 0.121 \pm 0.002$, obtained from LEP measurements at
the $Z$-peak \cite{PDG} 
and then run to two loops to $M_W=(80.41\pm0.10)$ GeV, 
$\alpha_s(m_b) = 0.232 \pm 0.003$ obtained from  QCD sum rules
in the $\Upsilon$ system \cite{upsilon}, 
$\alpha_s(M_\tau)=0.334 \pm 0.006$ \cite{ALEPH}, we then run this
value to two-loops up to $\overline m_c(m_c)=(1.23\pm0.05)$ GeV
in the $\overline{MS}$ scheme
\cite{NarisonMc} and get $\alpha_s(m_c)=0.42 \pm 0.01$.
In our approach \cite{BurasReviews},
 the penguin operators only get generated  from the charm
quark mass down  since the very small part due to the top quark is not 
relevant here. We have used the exact solutions of the renormalization 
group for the running of $\alpha_s$ and the quark masses.
For $C_3$, $C_4$, $C_5$, and $C_6$ we set the small imaginary part
due to the top loop to zero.
The results to one-loop accuracy are in Table \ref{WilsonS=1one-loop}
and for two-loops are in Table \ref{WilsonS=1two-loops}.
In the two-loop case we are using the NDR scheme.

If we run the short-distance contribution to two-loops, the matching
condition in (\ref{match4})  sets a further coefficient which
rends the matrix element scheme independent, i.e.
\ba
\label{r1NDR}
1+ \frac{\alpha_s(\mu)}{\pi} \, r_1 \, . 
\ea
In the case of the $\Delta S=2$ operator, the full two-loop
calculation can be found in \cite{S=2twoloops}. In the NDR scheme 
we have $r_1 = -7/6$ for the $\Delta S=2$ operator.
We obtained  it from the right eigenvalues of $\hat r_1^T$
 in \cite{BurasReviews}.
So we define
\be
\label{defBKhat}
\hat B_K  = \left(1+
\frac{\alpha_s(\mu)}{\pi}
\left[r_1+\frac{\gamma_2}{\beta_1}-\frac{\beta_2 \gamma_1}{\beta_1^2}
\right] \right) \, 
\left[\alpha_s(\mu)\right]^{\gamma_1/\beta_1}
B_K(\mu)\,.
\ee
With $\beta_1=-9/2$, $\beta_2=-8$, $\gamma_1=1$, and $\gamma_2=
-17/48$.  From the discussion in \cite{BurasReviews} and Section
\ref{Xboson} it can be seen that this definition is
scheme and renormalization scale independent.
We have shown in Table \ref{WilsonS=2} this factor
in front of $B_K(\mu)$ for the case of one-loop running 
labelled {\em One-loop}, two-loop running 
with $r_1=0$, labelled {\em Two-loops}, $r_1$ at its value, labelled
{\em Scheme-Independent (SI)},
 and a version where we use the exact solution
of the two-loop running with $r_1$ included so as to cancel the full
scheme-dependence there too, i.e.
\be
\label{defBKhatexp}
\hat B_K =\left(1+ \frac{\beta_2}{\beta_1}\frac{\alpha_s(\mu)}{\pi}
\right)^{[\gamma_2/\beta_2-\gamma_1/\beta_1 + 
(\beta_1/\beta_2) r_1] } \left[\alpha_s(\mu)\right]^{\gamma_1/\beta_1}
B_K(\mu)\,.
\ee
This is labelled {\em exp} in Table \ref{WilsonS=2}.
\begin{table}
\begin{center}
\begin{tabular}{|c|cccc|}
\hline
$\mu$(GeV)&One-Loop&Two-Loops&SI&{\em exp}\\
\hline
0.50& 1.04 &  1.11606 & 0.46894 & 0.63252 \\
0.60& 1.08 &  1.14653 & 0.76287 & 0.82690 \\
0.70& 1.12 &  1.18208 & 0.87769 & 0.91869 \\
0.80& 1.15 &  1.21045 & 0.94751 & 0.97817 \\
0.90& 1.17 &  1.23348 & 0.99680 & 1.02160 \\
1.00& 1.19 &  1.25267 & 1.03445 & 1.05546 \\
\hline
\end{tabular}
\end{center}\caption{\label{WilsonS=2} The coefficients to transform
$B_K(\mu)$ into $\hat B_K$. See text for an explanation of the different
columns. } 
\end{table}

The short-distance results for the $\Delta S=1$ Wilson
coefficients to two-loops and 
including the (\ref{r1NDR}) term which can be
found in the NDR scheme in \cite{two-loops,BurasReviews} 
for instance, are in Table \ref{WilsonS=1scheme}.
Here we give the one-loop results in Table \ref{WilsonS=1one-loop},
two-loop results with\footnote{Of course all quantities here are 
 matrices.}  $r_1=0$ at two-loops in Table \ref{WilsonS=1two-loops}
and the one with the scheme dependence properly removed, including $r_1$,
in Table \ref{WilsonS=1scheme}. It can be seen that the change from one to
two-loops in the NDR scheme 
is not so large but inclusion of the $r_1$ makes a large change.
\begin{table}
\begin{center}
\begin{tabular}{|c|cccccc|}
\hline
$\mu$(GeV)&$C_1$&$C_2$&$C_3$&$C_4$&$C_5$&$C_6$\\
\hline
0.50& -0.96466 & 1.59028&  0.01647& -0.03796&  0.01116& -0.04663\\
0.60& -0.84146 & 1.49560&  0.01067& -0.02626&  0.00801& -0.03037\\
0.70& -0.75899 & 1.43423&  0.00710& -0.01839&  0.00576& -0.02039\\
0.80& -0.69875 & 1.39058&  0.00468& -0.01263&  0.00403& -0.01356\\
0.90& -0.65222 & 1.35759&  0.00292& -0.00816&  0.00264& -0.00854\\
1.00& -0.61482 & 1.33159&  0.00158& -0.00455&  0.00149& -0.00467\\
\hline
\end{tabular}
\end{center}
\caption{\label{WilsonS=1one-loop} Wilson Coefficients of the operators
$Q_1$ to $Q_6$ at one-loop.} 
\end{table}
\begin{table}
\begin{center}
\begin{tabular}{|c|cccccc|}
\hline
$\mu$(GeV)&$C_1$&$C_2$&$C_3$&$C_4$&$C_5$&$C_6$\\
\hline
0.50& -0.80875&  1.48719&  0.13750& -0.26345&  0.01338& -0.27035\\
0.60& -0.74066&  1.43763&  0.05198& -0.11330&  0.02483& -0.09696\\
0.70& -0.65083&  1.36940&  0.03088& -0.07225&  0.02160& -0.05673\\
0.80& -0.58661&  1.32243&  0.02097& -0.05124&  0.01849& -0.03770\\
0.90& -0.53854&  1.28836&  0.01516& -0.03796&  0.01595& -0.02631\\
1.00& -0.50087&  1.26236&  0.01133& -0.02861&  0.01388& -0.01860\\
\hline
\end{tabular}
\end{center}
\caption{\label{WilsonS=1two-loops} Wilson Coefficients of the operators
$Q_1$ to $Q_6$ at two-loops in the NDR scheme.} 
\end{table}
\begin{table}
\begin{center}
\begin{tabular}{|c|cccccc|}
\hline
$\mu$(GeV)&$C_1$&$C_2$&$C_3$&$C_4$&$C_5$&$C_6$\\
\hline
0.50 &   -3.73959 &  4.02465 &  0.31282 & -0.43205  & 0.03267  &-0.33360\\ 
0.60 &   -1.89282 &  2.35657 &  0.10089 & -0.16996  & 0.02789  &-0.10140\\ 
0.70 &   -1.41708 &  1.95062 &  0.05666 & -0.10722  & 0.02258  &-0.05469\\
0.80 &   -1.17990 &  1.75588 &  0.03741 & -0.07706  & 0.01881  &-0.03390\\ 
0.90 &   -1.03270 &  1.63865 &  0.02665 & -0.05877  & 0.01597  &-0.02190\\
1.00 &   -0.93034 &  1.55917 &  0.01979 & -0.04625  & 0.01374  &-0.01397\\
\hline
\end{tabular}
\end{center}
\caption{\label{WilsonS=1scheme} Wilson Coefficients of the operators
$Q_1$ to $Q_6$ at two-loops, the NDR scheme dependence is 
removed as in discussed in Section \ref{Xboson}.} 
\end{table}

\end{document}